\documentclass[nopreprintline,3p]{elsarticle}
\usepackage{lineno,hyperref,siunitx,amssymb,amsmath}

\journal{Journal of Magnetism and Magnetic Materials}

\begin{document}

\begin{frontmatter}

\title{Helicity-independent all-optical switching of magnetization in ferrimagnetic alloys}

\author[address1]{C.~S.~Davies \corref{corr}}
\ead{carl.davies@ru.nl}
\cortext[corr]{Corresponding author}
\author[address2]{J.~H.~Mentink}
\author[address2]{A.~V.~Kimel}
\author[address2]{Th.~Rasing}
\author[address1]{A.~Kirilyuk}

\address[address1]{FELIX Laboratory, Radboud University, Toernooiveld 7, 6525 ED Nijmegen, The Netherlands}
\address[address2]{Radboud University, Institute for Molecules and Materials, 135 Heyendaalseweg, 6525 AJ Nijmegen, The Netherlands}

\begin{abstract}
We review and discuss the process of single-shot helicity-independent all-optical switching of magnetization by which a single suitably-ultrafast excitation, under the right conditions, toggles magnetization from one stable state to another. For almost a decade, this phenomenon was only consistently observed in specific rare-earth-transition-metal ferrimagnetic alloys of GdFeCo, but breakthrough experiments in recent years have revealed that the same behavior can be achieved in a wide range of multi-sublattice magnets including TbCo alloys doped with minute amounts of Gd, Gd/Co and Tb/Co synthetic ferrimagnets, and the rare-earth-free Heusler alloy Mn$_2$Ru$_x$Ga. Aiming to resolve the conditions that allow switching, a series of experiments have shown that the process in the ferrimagnetic alloys GdFeCo and Mn$_2$Ru$_x$Ga is highly sensitive to the pulse duration, starting temperature and the alloy composition. We argue here that the switching displayed by these two very different ferrimagnetic alloys can be generally understood within a single phenomenological framework describing the flow of angular momentum between the constituent sublattices and from the sublattices to the environment. The conditions that facilitate switching stem from the properties of these channels of angular momentum flow in combination with the size of the angular momentum reservoirs. We conclude with providing an outlook in this vibrant research field, with emphasis on the outstanding open questions pertaining to the underlying physics along with noting the advances in exploiting this switching process in technological applications.
\end{abstract}

\end{frontmatter}


\section{Introduction}\label{introduction}

The possibility to store information in the binary orientation of non-volatile magnetic bits has fueled, alongside the development of computing machinery, the emergence of a multi-billion-dollar industry aimed at developing ever smaller and faster data-storage technologies. The most popular conventional method of switching magnetization in a hard-disk-drive involves the application of a local magnetic field by a recording head that precessionally drives magnetization across the barrier posed by magnetocrystalline anisotropy. However, the magnetic trilemma -- whereby the constant downscaling of magnetic bits leads to stronger coercive fields that must be localized ever tighter -- has essentially obstructed any further progress in improving the specifications of conventional magnetic recording technologies~\cite{bean1959superparamagnetism,richter2007transition}, particularly in terms of writing speeds, storage densities and energy-efficiency. This has motivated intense exploration of new methods facilitating magnetic switching, with heat- and microwave-assisted magnetic recording being the flagship approaches currently adopted in state-of-the-art hard-disk-drives~\cite{kryder2008heat,challener2009heat,zhu2007microwave,okamoto2015microwave}.

Even until the 1990s, it was commonly assumed that the laws of thermodynamics adequately described the interaction between light and spins. This belief was upended by the seminal discovery of Bigot and Beaurepaire \textit{et al.} in 1996 that found that femtosecond laser pulses can destroy magnetization in ferromagnetic nickel on sub-picosecond timescales~\cite{Beaurepaire1996}. This result gave birth to the research field of ultrafast magnetism, devoted to learning how to manipulate magnetic materials on non-equilibrium timescales. Just over a decade later, Stanciu \textit{et al.}~\cite{Stanciu2007} observed that circularly-polarized femtosecond laser pulses are capable not only of destroying magnetization in gadolinium-iron-cobalt alloys (hereafter referred to as GdFeCo), but could actually reverse it without the help of any external magnetic fields. In 2012, it was even established that the circular polarization could be disregarded completely, leading to the process of helicity-independent all-optical switching~\cite{Ostler2012,Mentink2012}. These seemingly paradoxical results – disproving the long-held notion that magnetic fields must be at least somewhat involved to reverse magnetization - has since triggered intense research efforts~\cite{kirilyuk2013laser} aimed at engineering ever-faster switching capabilities under non-equilibrium conditions in a wide array of materials, ultimately targeting practical application within ultrafast memory technologies~\cite{el2020progress,polley2022progress}.

In this review, we briefly summarize recent measurements and our current understanding of helicity-independent all-optical switching of magnetization. First, we provide a historical introduction of the two pivotal experiments revealing both the feasibility and peculiar dynamics of the switching in GdFeCo alloys. Second, we summarize (in our view) the most important conditions that have been experimentally identified as governing the switching process in GdFeCo and the Heusler alloy Mn$_2$Ru$_x$Ga. Third, we present a phenomenological framework that allows us to broadly understand why the switching occurs in these two materials, and why it is governed by certain experimentally-observed restrictions. Fourth and finally, we provide an outlook aiming to draw attention to outstanding questions that remain in describing the underlying mechanism and manifestation of switching in other systems.

\section{All-optical switching of magnetization: An experimental overview}

\subsection{Different types of all-optical switching of magnetization}\label{DifferentTypes}

The umbrella term ``all-optical switching (AOS) of magnetization'' can refer to several distinct processes differentiated by the type of stimulus required and the switching that is displayed. In specific highly-absorbing media with uniaxial magnetocrystalline anisotropy, AOS can be achieved using either a single laser pulse of arbitrary polarization, whereby the magnetization of a ferrimagnet is toggled between two states~\cite{Ostler2012,Mentink2012}, or using multiple circularly-polarized laser pulses, which steers the one-way direction of switching as prescribed by the optical helicity~\cite{lambert2014all,mangin2014engineered,medapalli2017multiscale,Yamada2022efficient}. Alternatively, non-absorbing dielectrics also display resonance-driven AOS, importantly in the entire absence of heating. By pumping specific electronic transitions at resonance, one can precessionally switch magnetization with the direction of reversal controlled by the orientation of the linearly-polarized pulse's electric field~\cite{stupakiewicz2017ultrafast,stupakiewicz2019selection,frej2021all}. Alternatively, the selective excitation of a longitudinal optical phonon at resonance enables magnetization to be permanently switched via transient strain of the lattice~\cite{stupakiewicz2021ultrafast}. In this review, we restrict ourselves to discussing the process of single-shot helicity-independent all-optical switching found in metallic ferrimagnets.

\subsection{First measurements of helicity-independent all-optical magnetic switching}\label{FirstMeasurements}

\begin{figure}[h]
    \centering
    \includegraphics[width=90mm]{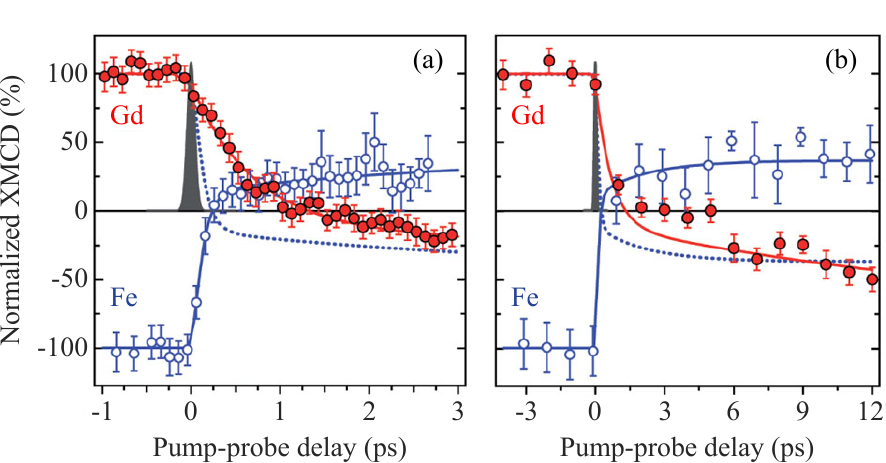}
    \caption{(a)-(b) Transient dynamics of the Fe (open circles) and Gd (filled circles) magnetic moments measured within the first \SI{3}{\pico\second} and \SI{12}{\pico\second} respectively. Error bars of the experimental data represent the statistical standard error. The measurements were performed at a sample temperature of \SI{83}{\kelvin} for an incident laser fluence of \SI{4.4}{\milli\joule\per\centi\meter^2}. The solid lines are fits according to a double exponential fit function. To facilitate comparison, the dashed line in both panels depicts the magnetization of the Fe sublattice inverted in sign. Adapted with permission from Ref.~\cite{Radu2011}.}
    \label{Radu_Image}
\end{figure}

Helicity-independent all-optical switching (HI-AOS) of magnetization refers to how a single optical pulse toggles the direction of magnetic polarity in a system with uniaxial anisotropy. The possibility of achieving such behavior was first suggested by Radu \textit{et al.}~\cite{Radu2011} with a study of the dynamics of ultrafast demagnetization of a Gd$_{25}$Fe$_{65.6}$Co$_{9.4}$ film. Using time-resolved x-ray magnetic circular dichroism (XMCD) measurements, Radu \textit{et al.} distinguished the magnetization of the antiferromagnetically-coupled Gd and Fe sublattices. The breakthrough result reproduced in Fig.~\ref{Radu_Image} reveals that the \SI{60}{\femto\second}-long pump pulse demagnetizes the Fe sublattice four times faster than the Gd sublattice. Moreover, the magnetization of the Fe sublattice actually crosses zero within \SI{300}{\femto\second} and subsequently reforms parallel to the still-demagnetizing Gd sublattice, giving rise to a transient ``ferromagnetic-like'' state. This state is absolutely forbidden to exist in equilibrium due to the strong antiferromagnetic coupling between the two magnetic sublattices. Ferromagnetic-like alignment persists for a further \SI{1.2}{\pico\second}, after which the magnetization of the Gd sublattice also crosses zero, with the sublattices then having antiparallel magnetic polarity in compliance with their antiferromagnetic coupling.

\begin{figure}[h]
    \centering
    \includegraphics[width=90mm]{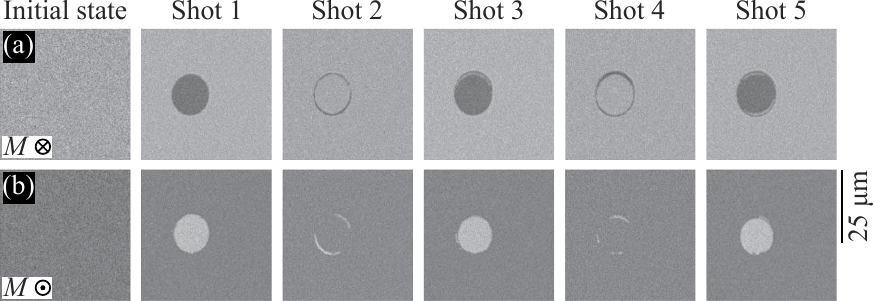}
    \caption{Typical magneto-optical images taken after exposing a Gd$_{24}$Fe$_{66.5}$Co$_{9.5}$ film to consecutive \SI{100}{\femto\second}-long optical pulses at room temperature. Each pulse has a fluence of \SI{2.30}{\milli\joule\per\centi\meter^2}. The images shown in rows (a)-(b) have different initial polarities of magnetization $M$ as indicated. Adapted with permission from Ref.~\cite{Ostler2012}.}
    \label{OstlerImage}
\end{figure}

Following the observation of the non-equilibrium ferromagnetic-like state, the process of HI-AOS was discovered~\cite{Ostler2012}. In these experiments, Ostler \textit{et al.} focused a single optical pulse, with wavelength \SI{800}{\nano\meter} and duration \SI{100}{\femto\second}, on to the surface of a Gd$_{24}$Fe$_{66.5}$Co$_{9.5}$ film at room temperature. The perpendicular magnetic anisotropy of GdFeCo alloys allows magnetization to point either parallel or antiparallel to the sample normal, allowing magneto-optical microscopy to conveniently detect magnetic domains with binary contrast. The images shown in Fig.~\ref{OstlerImage}, taken after exposing the GdFeCo film to consecutive optical pulses, clearly and irrefutably show that each pulse switches the magnetization deterministically. The lack of dependence on the optical helicity confirms that the switching, dynamically evolving in the manner shown in Fig.~\ref{Radu_Image}, is driven by the ultrashort thermal load delivered by the pulse.

Since these two landmark experiments, HI-AOS has been unambiguously identified in ferrimagnetic alloys of GdCo~\cite{el2019ultrafast}, (Gd,Tb)Co,~\cite{ceballos2021role,zhang2022role} and the Heusler alloys Mn$_2$Ru$_x$Ga (hereafter referred to as MRG)~\cite{banerjee2020single,davies2020exchange}. The discovery of such switching in ferrimagnetic alloys also led to predictions that synthetic ferrimagnets could exhibit similar behavior~\cite{evans2014ultrafast,gerlach2017modeling}, which has been identified thus far in multilayered nanostructures of Gd/FeCo~\cite{tsema2017laser}, Gd/Co~\cite{lalieu2017deterministic,beens2019comparing} and Tb/Co~\cite{aviles2019integration,aviles2020single}. Moreover, it is possible to supplement HI-AOS in GdFeCo with directionality of reversal using circularly-polarized pulses. This originates from preferential optical absorption via MCD, which gives rise to a slight mismatch in energy absorption between left- and right-handed circularly-polarized light. These pulses thus become capable of switching specific magnetic domains~\cite{khorsand2012role}. We emphasize that this is completely distinct from the other form of switching found in alternative ferri- and ferro-magnets, in which many circularly-polarized optical pulses, of left- or right-handed helicity, steer magnetic switching from an initially-demagnetized state~\cite{lambert2014all,mangin2014engineered,gorchon2016model,medapalli2017multiscale,Yamada2022efficient}.

\subsection{Experimental conditions for HI-AOS in GdFeCo and MRG}\label{Conditions}

In the following section, we present (in our view) the most important experimental conditions that provide for HI-AOS in GdFeCo and MRG. Both of these materials feature relatively strong inter-sublattice and dissimilar intra-sublattice exchange couplings. The magnetic moments of Gd and Fe differ by a factor of about four~\cite{Radu2011}. In contrast, the two sublattices of MRG correspond to Mn atoms at Wyckoff positions 4a and 4c, and so the sublattice-specific magnetic moments are rather similar with a ratio $\approx$1:1.2~\cite{betto2015site,vzic2016designing}.

\begin{figure}[h]
    \centering
    \includegraphics[width=90mm]{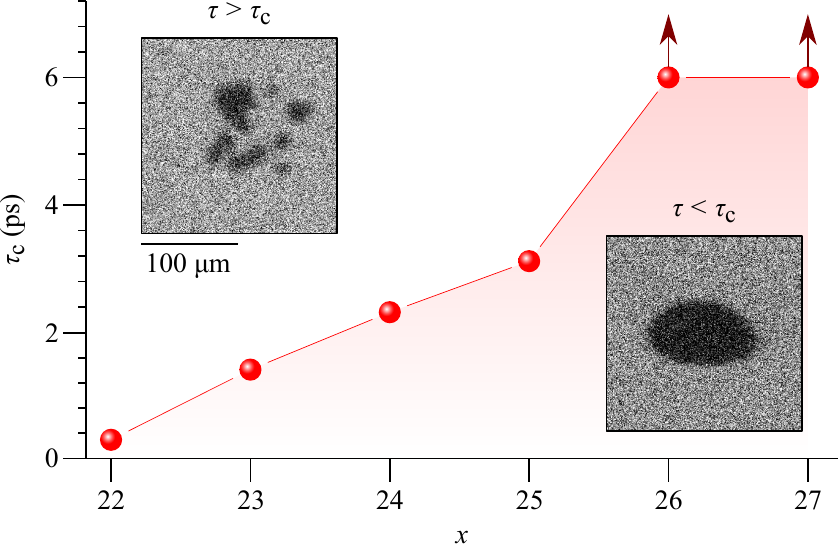}
    \caption{Critical pulse-duration threshold $\tau_{c}$ measured as a function of composition for Gd$_{x}$(FeCo)$_{100-x}$ alloys. HI-AOS is achieved if the pulse duration $\tau$ satisfies the condition $\tau<\tau_{c}$, but HI-AOS fails if $\tau>\tau_{c}$. The maximum achievable $\tau$ was \SI{6}{\pico\second} in these experiments, implying that $\tau_{c}>\SI{6}{\pico\second}$ for $x\geq26$. Insets: typical background-corrected magneto-optical images obtained for Gd$_{23}$(FeCo)$_{77}$ indicative of HI-AOS (bottom-right inset, $\tau=\SI{1.4}{\pico\second}$) and demagnetization (top-left inset, $\tau=\SI{1.5}{\pico\second}$). Adapted with permission from Ref.~\cite{davies2020pathways}.}
    \label{Davies2020_tau}
\end{figure}

A critically important parameter for HI-AOS is the length $\tau$ of the excitation pulse. Initial experiments studying HI-AOS predominantly used ultrashort (\SI{100}{\femto\second}-long) optical pulses with wavelengths in the vicinity of \SI{800}{\nano\meter}, reflecting the output of amplified Ti:Sa lasers. The possibility of using longer optical pulses was first suggested by Steil \textit{et al.} as far back as 2011, in experiments where helicity-dependent all-optical switching in a Gd$_{26}$Fe$_{64.7}$Co$_{9.3}$ sample was achieved using \SI{13}{\pico\second}-long pulses~\cite{steil2011}. At the time, this result was mired in an overarching controversy pertaining to the switching's origin, but this measurement represents the first sign indicating that AOS does not necessarily demand ultrashort optical pulses. A year later, Vahaplar \textit{et al.} identified a pulse-duration threshold $\tau_{c}$ for a Gd$_{24}$Fe$_{66.5}$Co$_{9.5}$ sample~\cite{vahaplar2012all}, whereby pulses shorter (longer) than $\tau_{c}=\SI{1.7}{\pico\second}$ were able (unable) to activate helicity-dependent switching. In contrast, the switching could be achieved in a film of composition Gd$_{26}$Fe$_{64.7}$Co$_{9.3}$ using \SI{2.1}{\pico\second}-long pulses. In 2016, experiments by Gorchon \textit{et al.} sparked substantial interest in the use of long optical pulses~\cite{gorchon2016role}, particularly with the pivotal discovery that $\tau_{c}$ can vary from \SI{1.5}{\pico\second} for Gd$_{24}$Fe$_{66}$Co$_{10}$ to as long as \SI{15}{\pico\second} for Gd$_{26}$Fe$_{64}$Co$_{10}$. This dependence on alloy composition was further explored by Davies \textit{et al.}, with experiments finding that $\tau_{c}$ increases monotonically with the percentage of Gd in Gd$_{x}$(FeCo)$_{100-x}$ alloys~\cite{davies2020pathways}, as shown in Fig.~\ref{Davies2020_tau}. This behavior has also been reproduced in both phenomenological modelling and atomistic simulations of HI-AOS~\cite{davies2020pathways,jakobs2021unifying}.

\begin{figure}[h]
    \centering
    \includegraphics[width=90mm]{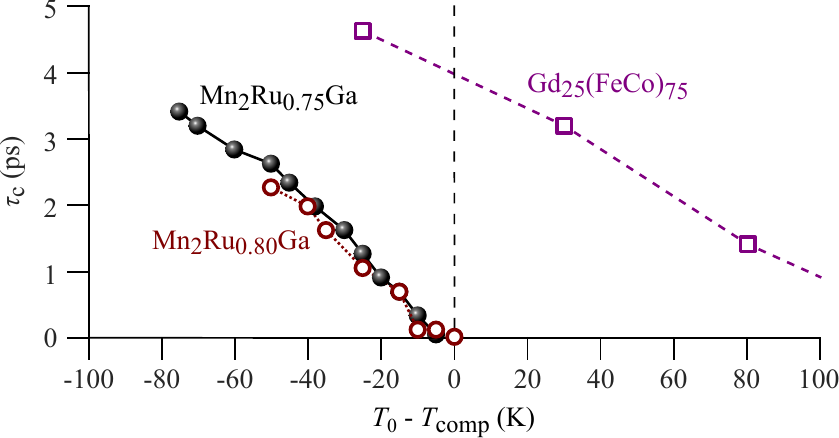}
    \caption{Threshold pulse duration $\tau_\textrm{c}$ for one GdFeCo and two MRG films measured as a function of the difference between the compensation temperature $T_\textrm{comp}$ and the measurement base temperature $T_\textrm{0}$. The compensation temperatures of Gd$_{25}$(FeCo)$_{75}$, Mn$_{2}$Ru$_{0.8}$Ga, and Mn$_{2}$Ru$_{0.75}$Ga are \SI{320}{\kelvin}, \SI{345}{\kelvin}, and \SI{370}{\kelvin}, respectively. Adapted with permission from Ref.~\cite{davies2020exchange}.}
    \label{Davies2020_comp}
\end{figure}

The variation of alloy composition in Gd$_{x}$(FeCo)$_{100-x}$ is qualitatively similar in impact to considering a fixed alloy composition at different equilibrium temperatures $T_{0}$~\cite{ostler2011crystallographically}. At a starting temperature $T=T_{0}$, the constituent sublattices of GdFeCo have angular momenta $S_\textrm{Gd}|_{T_{0}}$ and $S_\textrm{Fe}|_{T_{0}}$, hereafter labelled $S_\textrm{Gd,0}$ and $S_\textrm{Fe,0}$ respectively. Varying either $x$ or $T_{0}$ leads to a change in $S_\textrm{Gd,0}$ and $S_\textrm{Fe,0}$, naturally leading to the question of how the starting temperature $T_{0}$ influences HI-AOS. In the case of GdFeCo, Davies \textit{et al.} showed that increasing $T_{0}$ results in a monotonic decrease of $\tau_\textrm{c}$~\cite{davies2020exchange}. Importantly, this measurement showed that HI-AOS can be achieved at temperatures below and far above the compensation point $T_\textrm{comp}$ i.e. the temperature at which the net magnetization and angular momentum is zero. In a similar manner, the threshold $\tau_\textrm{c}$ for HI-AOS displayed by two samples of MRG decreases with $T_{0}$ (Fig.~\ref{Davies2020_comp})~\cite{davies2020exchange}. On the contrary, the HI-AOS tested in a total of fifteen MRG samples appears to always be constrained to starting temperatures below $T_\textrm{comp}$~\cite{banerjee2020single,davies2020exchange}.

The energy required for activating HI-AOS represents another important consideration. Laser-delivered optical pulses are almost always Gaussian in spatial distribution of energy, leading to a Gaussian distribution of heating at the focus. The non-uniform heating of the illuminated spot results in substantially more energy being deposited at the center of the spot compared to that deposited at the outer perimeter. This result is very useful for distinguishing the effects of pulse duration and energy. If the pulse duration is sufficient for switching but the absorbed fluence is excessive, the pulse generates a ring of switched magnetization encircling a spot of demagnetization~\cite{davies2020exchange,khorsand2012role,gorchon2016role}. At the same time, the Gaussian distribution conveniently allows absorbed-fluence thresholds associated with HI-AOS to be identified, with a minimum absorbed fluence $F_{c}$ being necessary for HI-AOS. Measurements of $F_{c}$ for GdFeCo range from 0.75 to 
\SI{3.14}{\milli\joule\per\centi\meter^2}~\cite{vahaplar2012all}, with $F_{c}=\SI{0.82}{\milli\joule\per\centi\meter^2}$~\cite{gorchon2016role} for a Gd$_{24}$(FeCo)$_{76}$ sample using $\tau=\SI{55}{\femto\second}$-long pulses at room temperature. As expected, the starting temperature, sample composition and pulse duration all influence $F_{c}$, but it is quite remarkable that increasing the pulse duration up to \SI{15}{\pico\second} for Gd$_{26}$(FeCo)$_{74}$ only increases $F_{c}$ to $\approx$\SI{1.5}{\milli\joule\per\centi\meter^2}~\cite{gorchon2016role}. Measurements for HI-AOS in MRG alloys have shown that $F_{c}$ can be minimized by reducing $\tau$ and bringing $T_\textrm{0}$ below but increasingly closer to $T_\textrm{comp}$~\cite{davies2020exchange}. 

The final parameter of the stimulus we consider is the type of excitation. Yang \textit{et al.} were the first to show experimentally that laser pulses with high photon energy are not compulsory for switching, with electronic pulses generated by an Auston switch being equally sufficient~\cite{yang2017ultrafast}. This represented a technological breakthrough since it can be argued that picosecond-long electrical pulses can already be excited in integrated circuits~\cite{el2020progress}. Further evidence of the indifference of HI-AOS to the type of excitation was provided by Davies \textit{et al.}, who showed that single optical pulses with photon energies ranging from \SI{1.55}{\electronvolt} to as low as \SI{50}{\milli\electronvolt} are similarly capable of driving HI-AOS in both GdFeCo and MRG~\cite{davies2020exchange,davies2020pathways}. While HI-AOS clearly only requires a gentle excitation of the electronic population just above the Fermi level, the latter measurements confirmed that such an excitation must still be shorter in duration than $\tau_{c}$. 

\section{Phenomenological description of AOS}\label{PhenomenologicalDescription}

In this section, we present a general theoretical framework describing ultrafast spin dynamics in multisublattice magnets~\cite{Mentink2012,mentink2012magnetism,radu2015ultrafast}. This framework, containing longitudinal relaxation terms of both exchange and relativistic origin, accounts for how angular momentum flows from one sublattice to another with conservation of the total angular momentum, and with accelerated and independent flow of angular momentum from each sublattice to an external bath (i.e. the lattice). We argue here that this model, while being phenomenological and somewhat simple in approach, can be used to qualitatively understand how HI-AOS can both be achieved and constrained by the conditions described in Section~\ref{Conditions}. 

\subsection{Equations of motion for longitudinal spin dynamics}\label{Equations}

The basis of our theoretical framework is the description of spin dynamics for a two-sublattice magnetic structure provided by Onsager's relations~\cite{onsager1931reciprocal}. Iwata~\cite{iwata1983thermodynamical,iwata1986irreversible} and Baryakhtar~\cite{baryakhtar1984phenomenological,bar1985phenomenological,baryakhtar1998phenomenological} independently developed this approach in the mid-1980s, showing that Onsager's relations naturally yields dynamics of the macroscopic length of the magnetization. Moreover, when taking into account the symmetry of the exchange interaction in multisublattice systems, dynamics of the length of the magnetization belonging to each sublattice is possible, even when the total angular momentum is conserved. These dynamics evolve across the timescale relevant to the exchange interaction, where conventional transverse (precessional) dynamics of the angular momentum can be considered to be negligible. In the limit of considering longitudinal dynamics of the macroscopic angular momentum $S_\nu$ belonging to sublattice $\nu$, the equations of motion for two non-equivalent collinear sublattices can be written as
\begin{eqnarray}
dS_a/dt &=& \lambda_{a}H_a + \lambda_e(H_a-H_b), \label{e:m1dot}\\
dS_b/dt &=& \lambda_{b}H_b + \lambda_e(H_b-H_a).\label{e:m2dot}
\end{eqnarray}
The angular momentum $S_\nu$, which can be either positive or negative in sign, is related to the magnetization $M_\nu$ by the gyromagnetic ratio $\gamma_\nu$ via $S_\nu=M_\nu/\gamma_\nu$. The $\lambda$ terms indicate the relative strength of the relaxation pathways. The relativistic transfer of angular momentum between sublattice $\nu=a,b$ and the environment is described by $\lambda_\nu$, whereas $\lambda_e$ is of exchange origin and stems from spin-spin interactions, conserving the total angular momentum by allowing angular momentum to transfer between the sublattices. The effective magnetic fields $H_\nu$ acting on each sublattice will be discussed extensively in the following Section.

\subsection{Non-equilibrium free energy and effective magnetic fields}\label{EffectiveFields}

The equations of motion given by Eqs.~(\ref{e:m1dot})-(\ref{e:m2dot}) assume that the time-dependent perturbation is induced by an external magnetic field. In the phenomenological theory presented here, the external field is replaced by an \textit{effective} magnetic field, with the latter being derived from a free energy taking into account internal interactions. Such an approach is very useful since it does not alter the structure of the equations of motion themselves.

Landau and Lifshitz were the first to introduce the concept of an effective magnetic field~\cite{landau1935theory}, defining it as the functional derivative of the macroscopic free energy $H_\nu=-\delta F/\delta S_\nu$, where $F$ is the free energy and $S_\nu=\langle\sum_i s^\nu_i\rangle/V$ is the operator of the total angular momentum per unit volume $V$ of sublattice $\nu$. Aiming to derive the macroscopic free energy from a microscopic Hamiltonian $\mathcal{H}$, we use for simplicity the Heisenberg spin model, which can be written for 2 sublattices ($\nu=a,b$) in the form
\begin{equation}\label{e:heisenberg2}
 \mathcal{H} = -{\sum_{i \in a \atop i^\prime \in a}}^\prime J^{aa}_{ii^\prime}s_i^as_{i^\prime}^a - {\sum_{i \in b \atop i^\prime \in b}}^\prime  J^{bb}_{ii^\prime}s_i^bs_{i^\prime}^b - {\sum_{i \in a \atop i^\prime \in b}}^\prime  J^{ab}_{ii^\prime}s_i^as_{i^\prime}^b - {\sum_{i \in b \atop i^\prime \in a}}^\prime J^{ab}_{ii^\prime}s_i^bs_{i^\prime}^a,
\end{equation}
Here, $J_{ii^\prime}^{\nu\nu^\prime}$ are exchange parameters and $\sum_{i,i^\prime}^\prime=\sum_i\sum_{i^\prime\neq i}$ indicates a double summation where each sum runs over a whole sublattice. In the following, and similar in approach to atomistic spin dynamics, we treat the spins $s_i^\nu$ as classical, but we note that calculations for quantum spins also proceed in a similar way~\cite{abubrig2001}.

To calculate the relationship between $\mathcal{H}$ and the free energy $F$, it is possible in principle to use the fundamental relation $F(\mathcal{H})=\langle\mathcal{H}\rangle-T\mathbb{S}$
where $\mathbb{S}$ is the entropy and $T$ the temperature of the medium in which the spin system is embedded. While this relationship can be used both in and out of equilibrium, actual calculation is difficult, and so we limit ourselves to the mean field approximation $S_\nu=N_\nu\langle s^\nu_i\rangle$ where $N_\nu$ is the number of spins of sublattice $\nu$ per unit volume.
We can ensure that the the mean-field approximation of $F$ is still valid when considering non-equilibrium scenarios by performing statistical perturbation theory. Writing $\mathcal{H}=H_0+(\mathcal{H}-H_0)$, we obtain (to first order in $\mathcal{H}-H_0$)
\begin{equation}
F \leq F(H_0) + \langle\mathcal{H}-H_0\rangle_0\equiv\Phi,
\end{equation}
where $\langle x \rangle_0=\langle x \exp(-\beta H_0)\rangle/\langle \exp(-\beta H_0)\rangle$ indicates averaging over the equilibrium distribution function of the trial Hamiltonian $H_0$. In the classical and quantum case, this inequality is named Gibbs and Bogoliubov respectively~\cite{falk1970inequalities}.

To arrive at the mean-field approximation, we choose the simple form
\begin{equation}
H_0=-\sum_{i\in a}h_as_i^a -\sum_{i\in b}h_bs_i^b,
\end{equation}
where $h_\nu$ are variational parameters. Direct calculation gives
\begin{equation}
F(H_0)=-\beta^{-1}\left[N_a\ln Z_a + N_b \ln Z_b\right],
\end{equation}
\begin{equation}
\langle H_0 \rangle_0=-\frac{\partial}{\partial\beta}(N_a \ln Z_a +N_b \ln Z_b)=-N_ah_as_a-N_ah_bs_b,
\end{equation}
and
\begin{equation}
\langle \mathcal{H} \rangle_0= -N_a z_{aa} J_{aa}s_a^2 - N_b z_{bb} J_{bb}s_b^2 -2N_pJ_{ab} s_a s_b,
\end{equation}
with
\begin{equation}\label{e:smf}
s_\nu=\langle s_i^\nu \rangle_0=\frac{\partial \ln Z_\nu}{\partial \beta h_\nu}=\sigma_\nu\mathcal{L}(\beta h_\nu \sigma_\nu).
\end{equation}
Here, we use the Langevin function $\mathcal{L}(x)=\coth(x)-1/x$ and the single-spin partition function $Z_\nu=4\pi\sinh(\beta h_\nu \sigma_\nu)/(\beta h_\nu \sigma_\nu)$, with $\sigma_\nu=|s^\nu_i|$ being the length of the local spin moment. $N_p=N_az_{ab}=N_bz_{ba}$ represents the number of pairs and $z_{\nu\nu^\prime}$ the number of neighbors in sublattice $\nu^\prime$ of a spin in sublattice $\nu$. Eq.~(\ref{e:smf}) is a one-to-one relation between the mean-field spin moment per site $s_\nu$ and the variational parameter $h_\nu$. Hence, we can consider $h_\nu$ to be an explicit function of $s_\nu$, and thereby write the mean-field free energy both in and out of equilibrium in the form
\begin{equation}\label{e:Fmffull}
\Phi(s_a,s_b)=N_af_a(s_a)+N_bf_b(s_b)-2N_pJ_{ab}s_as_b,
\end{equation}
with
\begin{equation}\label{e:fmfnu}
f_\nu(s_\nu)=-\left[ \beta^{-1} (\ln Z_\nu - \eta_\nu s_\nu/\sigma_\nu) + z_{\nu\nu} J_{\nu\nu}s_\nu^2 \right],
\end{equation}
where $\eta_\nu=\beta h_\nu \sigma_\nu = \mathcal{L}^{-1}(s_\nu/\sigma_\nu)$. Equations~(\ref{e:Fmffull})-(\ref{e:fmfnu}) provide the link between the microscopic spin Hamiltonian and the macroscopic non-equilibrium free energy in the mean-field approximation.

We are now in a position to explicitly derive the effective fields from the free energy. From Eq.~(\ref{e:smf}), we find that
\begin{equation}
\frac{\partial}{\partial s_\nu}(\ln Z_\nu - \eta_\nu s_\nu/\sigma_\nu) = \frac{\partial \ln Z_\nu}{\partial\eta_\nu}\frac{\partial\eta_\nu}{\partial s_\nu} - \frac{\partial\eta_\nu}{\partial s_\nu}\frac{s_\nu}{\sigma_\nu} - \eta_\nu/\sigma_\nu = - \eta_\nu/\sigma_\nu,
\end{equation}
such that the effective fields $H_\nu=-\frac{1}{N_\nu}\frac{\partial\Phi}{\partial s_\nu}$ become
\begin{eqnarray}
\label{e:h1eff}
 H_a = -\beta^{-1}\eta_a/\sigma_a + 2 z_{aa} J_{aa}s_a + 2z_{ab}J_{ab} s_b,\\
 \label{e:h2eff}
 H_b = -\beta^{-1}\eta_b/\sigma_b + 2 z_{bb} J_{bb}s_b + 2z_{ba}J_{ab} s_a,
\end{eqnarray}

In equilibrium, and with the assumption that no external fields are present, the effective fields vanish. We thus obtain the special values $\overline{h}_\nu=\beta^{-1}\overline{\eta}_\nu/\sigma_\nu$ given by
\begin{eqnarray}
\label{e:h1eq}
\overline{h}_a= 2 z_{aa} J_{aa}\overline{s}_a + 2z_{ab}J_{ab} \overline{s}_b,\\
\label{e:h2eq}
\overline{h}_b= 2 z_{bb} J_{bb}\overline{s}_b + 2z_{ba}J_{ab} \overline{s}_a,
\end{eqnarray}
where the equilibrium values $\overline{s}_\nu$ can be determined by the self-consistent solution of the coupled set of equations
\begin{equation}\label{e:smfequil}
\overline{s}_\nu=\sigma_\nu\mathcal{L}(\beta \sigma_\nu\overline{h}_\nu).
\end{equation}
The same result is obtained in the usual equilibrium mean-field theory with the quantities $\overline{h}_\nu$ being interpretable as Weiss fields. Such equilibrium mean-field theory is often used in the derivation of effective fields entering precessional dynamics. However, as long as the magnetic sublattices are not yet in equilibrium with each other, $h_\nu\neq\overline{h}_\nu$. It is therefore mandatory to use $h_\nu$ in the effective fields to obtain the correct relaxation to equilibrium.

\subsection{Classification of dynamics}\label{ClassificationDynamics}

Equations~(\ref{e:h1eff})-(\ref{e:h2eff}), in combination with Eqs.~(\ref{e:m1dot})-(\ref{e:m2dot}) form a closed set of equations which provide the basis for phenomenologically describing the non-equilibrium dynamics of the longitudinal angular momenta of multi-sublattice magnets. Generally, the exchange energies $f_\nu$ and $J_{ab}$ in $F$ are parametrically dependent on the temperature of the environment. Hence, the aforementioned set of equations can be used to simulate the dynamics in response to heat pulses. Such numerical simulations were employed in Ref.~\cite{davies2020pathways} and it was found that the model encompasses the counterintuitive switching of magnetic sublattices in various regimes, featuring all known results from much more computationally-demanding atomistic simulations~\cite{Ostler2012,atxitia2014controlling}. Here, we use this result to classify the dynamics in two distinct regimes~\cite{Mentink2012,davies2020pathways}. For this purpose, it is convenient to analyze the dynamics by expanding the free energy to leading order in $s^2_\nu-\overline{s}_\nu^2$, which recovers the familiar phenomenological Landau expressions~\cite{mentink2012magnetism}.

The first regime is defined for $T\gg T_{\rm C}$, which we call the temperature-dominated regime. Such a scenario can be realized by suddenly heating the electron system using a fs laser pulse. On the timescale of 10-\SI{100}{\femto\second}, the electronic temperature, and hence the value of the ``temperature'' that enters the expression for the nonequilibrium free energy, far exceeds the equilibrium Curie temperature~\cite{Beaurepaire1996,steil2011}. The system thus behaves across this timescale like a paramagnet, and so we can write $f_\nu=S_\nu^2/(2\chi_\nu)$ where $\chi_\nu\sim1/T$ denotes the longitudinal susceptibility of sublattice $\nu$. Since, in the paramagnetic regime, $J_{ab}\ll k_BT$, the inter-sublattice interaction can be neglected, and the independent transfer of angular momentum from each sublattice to the environment dominates the dynamics. Consequently, the sublattices exhibit decoupled Bloch relaxation with a characteristic longitudinal relaxation time given by $\tau_\nu=\chi_\nu/\lambda_\nu$. Microscopic calculations \cite{brown1963thermal,kubo1970brownian,garanin1997fokker} show that this is equivalent to
\begin{equation}\label{e:taui}
\tau_\nu=\sigma_\nu/(2\alpha_\nu k_BT),
\end{equation}
where $\alpha_\nu$ is a microscopic parameter of relativistic origin determining the coupling to the heat bath. Importantly, Eq.~(\ref{e:taui}) shows that the atomic spin moment $\sigma_\nu$ controls the speed of the longitudinal relaxation, with smaller magnetic moments relaxing faster. This finding is in accordance with the result shown in Fig.~\ref{Radu_Image}, since the magnetic moment of iron is roughly four times smaller than that of gadolinium~\cite{ostler2011crystallographically}.

The second regime, hereafter-referred to as the exchange-dominated regime, is relevant in non-equilibrium scenarios where $T<T_{\rm C}$, typically appearing on the picosecond timescale. Exchange interactions in an ordered system are generally stronger than relativistic interactions and so the exchange-dominated regime is characterized by $\lambda_\nu\ll\lambda_e$. In this regime, the transfer of angular momentum from one sublattice to the other dominates, with the conservation of total angular momentum $S_{a}+S_{b}=constant$ in the limit of dynamics driven entirely by exchange-relaxation. As a consequence, for any form of the free energy $F$, the changes of the sublattice-specific angular momentum sum up (approximately) to zero, leading to $dS_a/dt=-dS_b/dt$. This yields highly counter-intuitive dynamics when the spin of one of the sublattices is close to zero, whereby $dS_\nu/dt$ remains finite even when $S_\nu=0$. This represents the only pathway by which the coupled spins can cross zero and reverse sign. 

\begin{figure}[h]
\center{\includegraphics[width=90mm]{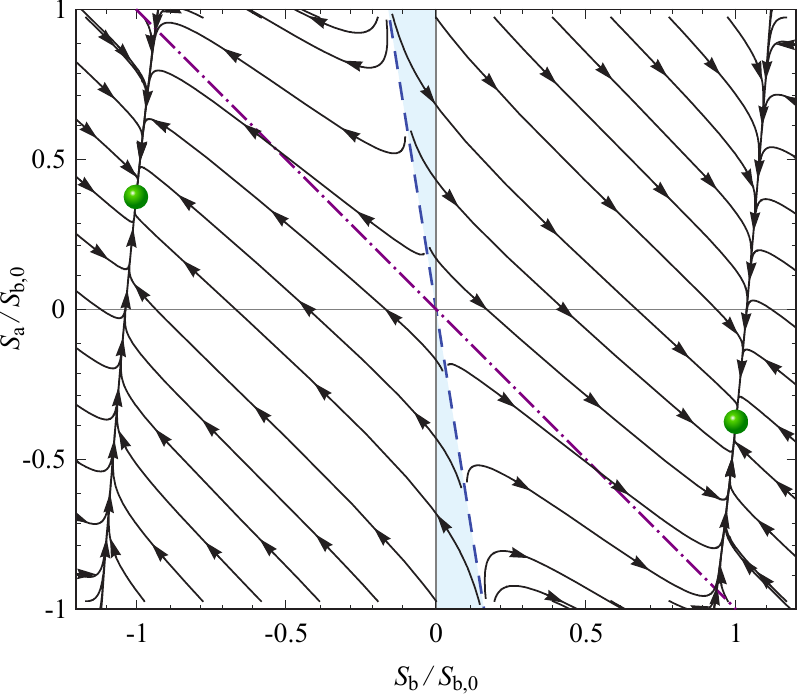}}%
\caption{Numerical solution of the longitudinal equations of motion in the exchange-dominated regime. The evolution of $S_a$ is shown as function of $S_b$ for various initial conditions, where both are normalized to the equilibrium value of angular momentum $S_{b,0}$. The green spheres indicate stable equilibrium points and the arrows indicate the direction of relaxation with increasing time. The origin is a saddle point, and the dashed blue line indicates a stable manifold of solutions whereas the dotted-dashed purple line corresponds to the condition $S_a+S_b=0$. The blue shaded area encompasses the initial conditions from which the longitudinal relaxation will proceed to reversal via temporal ferromagnetic alignment. Adapted with permission from Ref.~\cite{Mentink2012}.\label{f:sim1}}
\end{figure}

To illustrate the effect of exchange-dominated dynamics schematically in the simplest possible way i.e., without considering specific pulse and system parameters, we can solve Eqns.~(\ref{e:m1dot})-(\ref{e:m2dot}) at fixed temperature of the heat bath. For a typical rare-earth ($\nu=a$) transition-metal ($\nu=b$) ferrimagnet, $J_{aa}$ is substantially smaller than $J_{bb}$, and so one can model the sublattices's free energies in the form of $f_a=AS_a^2/2$ and $f_b=B(S_b^2-\overline{S}_b^2)^2/4$. Using exemplary values $A/B=0.4$, $B=1=\overline{S}_b$, $J_{ab}/B=-0.15$, $\lambda_a=\lambda_b=0.15$ and $\lambda_e=1$, Mentink \textit{et al.} calculated the results shown in Fig.~\ref{f:sim1}~\cite{Mentink2012}. This phase diagram conveniently visualizes the dynamics of $S_a$ as a function of $S_b$ for various initial conditions in one graph. In this phase plane, a pure form of exchange-relaxation appears as trajectories inclined at -$45^{\circ}$ relative to the horizontal axis (parallel to the dashed-dotted purple line), fulfilling $dS_a/dt=-dS_b/dt$.

In general, the phase diagram shown in Fig.~\ref{f:sim1} shows the trajectory of longitudinal angular-momenta dynamics one would obtain for a ferrimagnetic system that is brought out of equilibrium. The starting situation of the ferrimagnet in stable equilibrium is indicated by the green spheres in the upper-left or lower-right quadrants i.e., the quadrants in which $S_a$ and $S_b$ have opposite sign. Following excitation, the ferrimagnet is brought out of equilibrium, bringing the system to a different ``starting'' coordinate ($S_b$,$S_a$) on the phase diagram. The solid lines with arrows show the ferrimagnet's direction of subsequent relaxation to equilibrium. For a system starting in the top-left quadrant, the unshaded majority of the phase space leads to a return back to its original state. If, however, the excitation substantially and sufficiently demagnetizes $S_b$ with $|S_b|<S_a$, the ferrimagnet is driven in to the blue shaded manifold. This is a special region where the angular momentum of each sublattice will move in to the top-right quandrant, corresponding to the ferromagnetic-like state detected experimentally in Fig.~\ref{Radu_Image}. From this quadrant, HI-AOS follows.

Going further, we can express the requirement for sublattice reversal in the blue manifold to occur when
\begin{equation}\label{e:criteriontfm}
\left. \frac{dS_b}{dt} \right\rvert_{S_b=0}>0 \Leftrightarrow 2\frac{\partial f_a}{\partial S_a^2}>-J_{ab}(1+\lambda_b/\lambda_e)>0.
\end{equation}
Substitution of the Landau form of $f_a$ given above in Eq.~(\ref{e:criteriontfm}) yields $\lambda_b<\lambda_e(A/|J_{ab}|-1)$. Since by definition $\lambda_b\geq0$, we find that reversal is thus only possible when exchange relaxation is included ($\lambda_e>0$). 

The phase diagram shown in Fig.~\ref{f:sim1} represents a powerful tool for understanding how exactly HI-AOS can arise. In principle, any excitation which brings the angular momenta of the two sublattices from the stable equilibrium point to the blue shaded manifold will result in HI-AOS. We emphasize here that this result shows that the transient ferromagnetic state is \textit{not} the critical prerequisite for the switching, but rather it is the strongly demagnetized state indicated by the blue shaded manifold, in which $|S_b|<S_a$ and $S_b$ is substantially demagnetized. By considering different system parameters, such as inter- and intra-sublattice exchange couplings, starting temperature and relaxation constants, the blue manifold shown in Fig.~\ref{f:sim1} can widen, shrink and even rotate. We refer the reader to Ref.~\cite{mentink2012magnetism} for examples of how such behavior can be realized.

\subsection{Phenomenological framework of HI-AOS in GdFeCo and MRG}\label{FrameworkExamples}

The main result of Section~\ref{ClassificationDynamics} is that HI-AOS can only be achieved by propelling the ferrimagnetic system in to the strongly non-equilibrium state shaded in blue in Fig.~\ref{f:sim1}. The question that we aim to answer, in this Section, relates to \textit{how} we can drive entry in to this non-equilibrium state. We emphasize that we do not aim to use the phenomenological model to quantitatively reproduce experimental measurements, since this would involve fitting many parameters that would not bring much further insight. A phenomenological approach rather offers substantial predictive powers due to its lack of material-specific assumptions and computational overheads. Indeed, we find that our phenomenological approach qualitatively explains the unique kinetics of HI-AOS and the origins of the experimentally-observed conditions described in Section~\ref{Conditions}.

\begin{figure}[t]
    \centering
    \includegraphics[width=90mm]{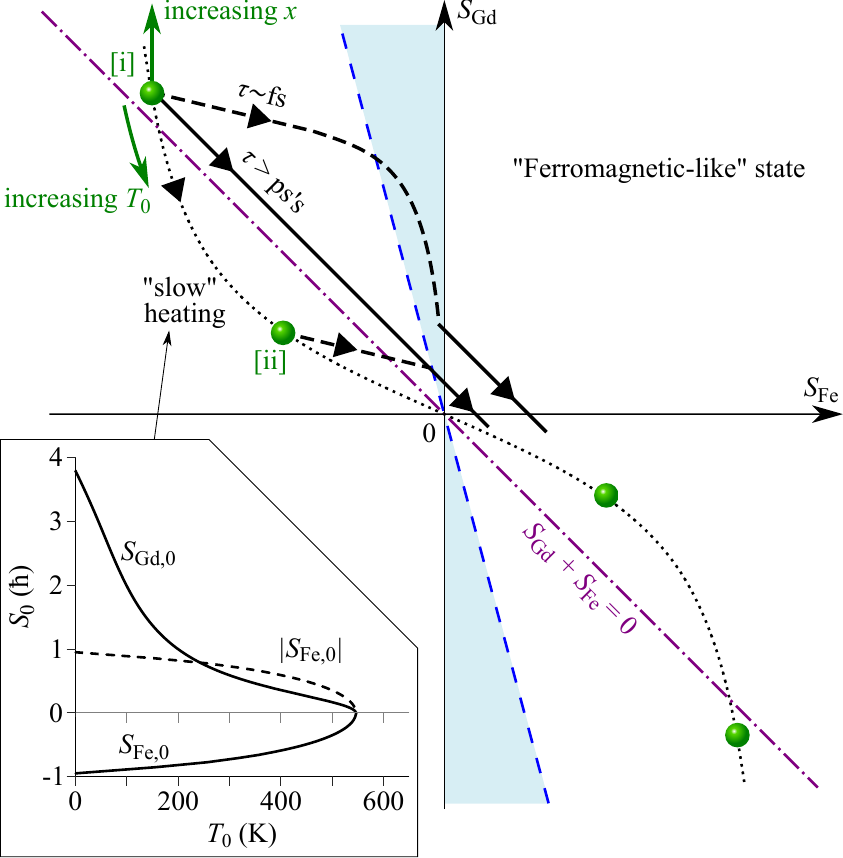}
    \caption{Conceptual phase diagram showing the different pathways for thermally-induced relaxation of the sublattice-resolved angular momentum $S_\textrm{Fe}$ and $S_\textrm{Gd}$ of the ferrimagnetic alloy Gd$_x$Fe$_{100-x}$. The green spheres in the top-left and bottom-right quadrants indicate example positions of equilibrium (labelled [i] and [ii]). By varying $x$ or the starting temperature $T_{0}$, the equilibrium states move across the phase diagram, with the dotted line corresponding to the scenario of ``slow'' heating in equilibrium. Excitation of the ferrimagnet by a thermal pulse of duration $\tau$ leads to different trajectories of demagnetization, with a femtosecond pulse activating decoupled Bloch relaxation (dashed black line) followed by exchange-relaxation (solid black line line), and with a longer pulse activating exchange-relaxation only (solid black line). After both $S_\textrm{Fe}$ and $S_\textrm{Gd}$ change sign, HI-AOS finalizes via ``slow'' relaxation to the equilibrium state in the bottom-right quadrant. The inset shows the thermal dependence of the equilibrium angular momentum $S_\textrm{Gd,0}$ and $S_\textrm{Fe,0}$ (solid lines). To facilitate comparison, we also show $|S_\textrm{Fe,0}|$ (dashed line). Adapted with permission from Ref.~\cite{davies2020pathways}.}
    \label{f:concept1}
\end{figure}

To model GdFeCo alloys in equilibrium, we can solve either Eqs.~(\ref{e:h1eq})-(\ref{e:h2eq}) self-consistently or Eqs.~(\ref{e:h1eff})-(\ref{e:h2eff}) with Eqs.~(\ref{e:m1dot})-(\ref{e:m2dot}) in equilibrium. These both yield the equilibrium temperature dependence of the angular momentum for each constituent sublattice within the ferrimagnet~\cite{Note1}, shown in the inset of Fig.~\ref{f:concept1}. In the spirit of using phase diagrams to explain the process of HI-AOS, we recast in the main panel of Fig.~\ref{f:concept1} the equilibrium thermal dependence as a dotted black line. This represents the size of the angular momentum reservoirs $S_\textrm{Fe}(T_{0})$ and $S_\textrm{Gd}(T_{0})$ at equilibrium. At a given starting temperature, the angular momentum of the ferrimagnet is confined to a point on this line (exemplary green spheres labelled [i] and [ii]). ``Slow'' variation of the temperature in equilibrium results in the angular momentum moving along this line only, with both sublattices experiencing the same temperature $T_{0}$.

Let us now consider the HI-AOS displayed by GdFeCo. As discussed in Section~\ref{ClassificationDynamics}, the longitudinal dynamics of angular momentum can be classified in two distinct regimes, either being temperature- or exchange-dominated. The first scenario corresponds to ultrafast femtosecond heating, which essentially decouples the two sublattices with loss of angular momentum to the environment at a rate inversely proportional to the sublattice's magnetic moment [Eq.~(\ref{e:taui})]. Because of the weakness of $J_{Gd-Fe}$ and the magnetic moment of Fe being four times smaller than that of Gd, the femtosecond pulse initially demagnetizes Fe roughly four times faster than Gd. This trajectory is sketched in Fig.~\ref{f:concept1} as the dashed black line, traversing the phase diagram with a gradient of $\approx$4:-1. With the iron sublattice demagnetizing more rapidly, $S_\textrm{Fe}$ approaches close to the axis $S_\textrm{Fe}=0$. The trajectory then bends towards the origin ($S_\textrm{Fe}=0,S_\textrm{Gd}=0$) since this relativistic relaxation can only remove angular momentum from the sublattices. With the electrons and lattice equilibrating across the timescale $\tau_{e-l}\approx\SI{2}{\pico\second}$~\cite{koopmans2010explaining}, exchange relaxation begins to dominate (solid black line)~\cite{bergeard2014ultrafast}. The exchange-dominated dynamics, conserving the total angular momentum of the two sublattices, drives the ferrimagnet in to the ferromagnetic-like state with the Fe sublattice receiving angular momentum from the Gd sublattice i.e., $dS_\textrm{Fe}/dt=-dS_\textrm{Gd}/dt$. In this case, the trajectory is given by $S_{a}+S_{b}=constant$, parallel to the dashed-dotted purple line. The exchange relaxation then continues to push $S_\textrm{Gd}$ across zero as well, resulting in $S_\textrm{Gd}$ and $S_\textrm{Fe}$ now having switched signs. The spins then equilibrate with the lattice across distinct timescales, and subsequent cooling eventually finalizes the HI-AOS with the ferrimagnet's angular momentum residing in the bottom-right quadrant.

A similar description can be used to explain the HI-AOS achieved by a substantially-stretched excitation, with duration on the order of several picoseconds. In this scenario, the spins do not experience an intense sharp change in ``temperature'' that brings about decoupled Bloch relaxation. Instead, the sublattices primarily demagnetize via exchange-relaxation, transferring angular momentum from the Gd sublattice to the Fe one. Conserving the total angular momentum of the system, the trajectory of demagnetization only follows the solid black line across Fig.~\ref{f:concept1}. Provided that the ferrimagnet's starting point is above the dashed-dotted purple line e.g. at position [i], such exchange-relaxation is still capable of driving the system in to the strongly non-equilibrium state (blue manifold in Fig.~\ref{f:concept1}) from which HI-AOS emerges. For HI-AOS to occur, $S_\textrm{Fe}$ must cross zero before $S_\textrm{Gd}$~\cite{beens2019comparing,davies2020pathways,graves2013nanoscale,atxitia2014controlling}. This condition becomes evident when considering the state in which both sublattices are considerably demagnetized and are starting to recover. The sublattice with the stronger intra-sublattice exchange interaction will recover faster, and so HI-AOS demands $S_\textrm{Fe}$ to cross zero first. It is also important during this process that both $S_\textrm{Fe}$ and $S_\textrm{Gd}$ do not simultaneously fall to too low a value where thermal spin correlations can dominate. This scenario leads to loss of magnetic memory and subsequent demagnetization, as encountered if the absorbed fluence is excessive.

The two pathways described above indicate the routes through which HI-AOS can be achieved. Equipped with this understanding, we can comprehend how HI-AOS depends on the sample composition $x$, as shown experimentally and numerically in Fig.~\ref{Davies2020_tau}. Increasing the percentage of Gd in the Gd$_{x}$(FeCo)$_{100-x}$ alloy rapidly increases the size of the equilibrium angular momentum reservoir $S_\textrm{Gd}$, pushing the initial equilibrium state upwards in the top-left quadrant in Fig.~\ref{f:concept1}. Thus, increasingly longer pulses become capable of driving the switching since the system can tolerate increasingly larger deviations from exchange-driven dynamics introduced by leakage of angular momentum to the lattice. Quantitative modelling of this scenario, using Eqs.~(\ref{e:m1dot})-(\ref{e:m2dot}) in combination with Eqs.~(\ref{e:h1eff})-(\ref{e:h2eff}), was presented by Davies \textit{et al.} in Ref.~\cite{davies2020pathways}.

Similarly, we can also understand why the compensation temperature is rather insignificant in the process of HI-AOS found in GdFeCo. At elevated temperatures $T_{0}>T_\textrm{comp}$, the ferrimagnet's starting point ($S_\textrm{Fe,0}$,$S_\textrm{Gd,0}$) sits below the dashed-dotted purple line (e.g. position [ii] in Fig.~\ref{f:concept1}), and so exchange-driven dynamics becomes incapable of driving the angular momenta in to the blue shaded manifold. Decoupled Bloch relaxation, however, can still meet this condition, thus explaining why only shorter pulses are capable of achieving HI-AOS at higher starting temperatures~\cite{davies2020pathways} or in alloys of Gd$_{x}$(FeCo)$_{100-x}$ with lower $x$ (Fig.~\ref{Davies2020_tau}).

\begin{figure}[h]
    \centering
    \includegraphics[width=90mm]{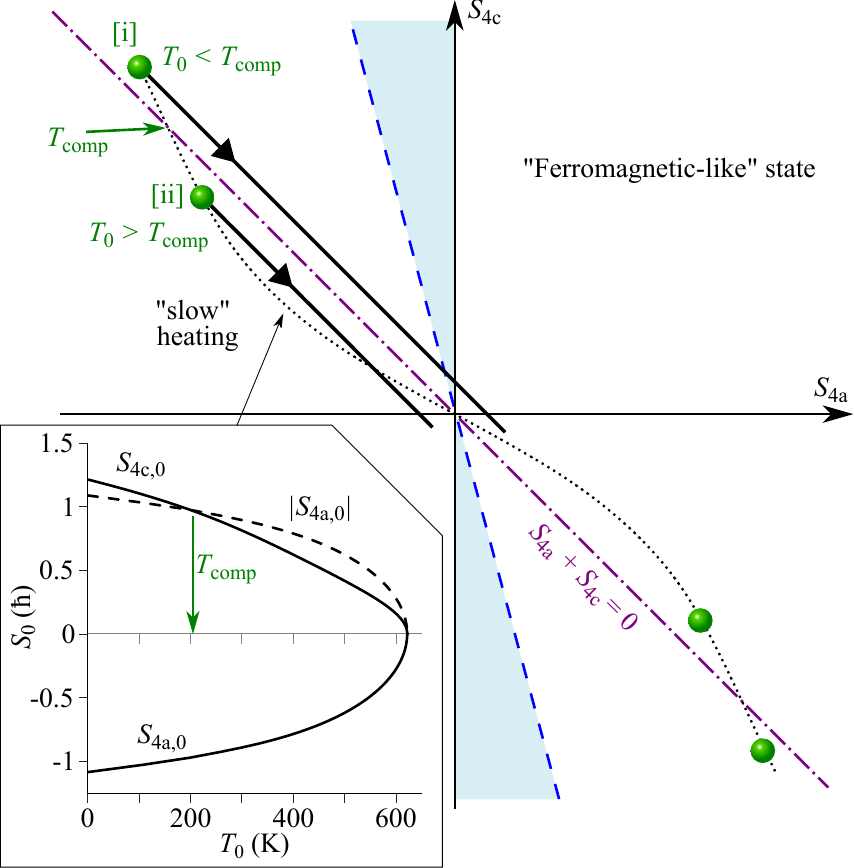}
    \caption{Conceptual phase diagram showing the different pathways for thermally-induced relaxation of the sublattice-resolved angular momenta $S_\textrm{4a}$ and $S_\textrm{4c}$ of the ferrimagnetic alloy Mn$_{2}$Ru$_x$Ga. The green spheres in the top-left and bottom-right quadrants indicate example positions of equilibrium. By varying the starting temperature $T_{0}$, the equilibrium states move across the phase diagram, lying above or below the compensation point $T_{comp}$ where $S_\textrm{4a}+S_\textrm{4c}=0$. Non-equilibrated excitation of the ferrimagnet primarily activates exchange relaxation (solid black line). HI-AOS is activated if $S_{4a}$ switches sign first, which can only be achieved when $T_{0}<T_{comp}$. If instead $T_{0}>T_{comp}$, $S_{4c}$ switches sign first and HI-AOS fails. The dotted line corresponds to the scenario of ``slow'' heating in equilibrium. The inset shows the thermal dependence of the equilibrium angular momentum $S_\textrm{4c,0}$ and $S_\textrm{4a,0}$ (solid lines). To facilitate comparison, we also show $|S_\textrm{4a,0}|$ (dashed line). Adapted with permission from Ref.~\cite{davies2020exchange}.}
    \label{f:concept2}
\end{figure}

We now turn to the case of MRG, with the equilibrium thermal dependence of $S_\textrm{4a}$ and $S_\textrm{4c}$ given in the inset of Fig.~\ref{f:concept2}. In this alloy, the sublattice-specific magnetic moments are similar (ratio $\approx$1:1.2) and are strongly coupled ($J_{4a-4c}=$\SI{-55}{\milli\electronvolt}). Thus, decoupled Bloch relaxation at sublattice-specific rates, as expected from a femtosecond excitation, does not substantially dominate on the timescale before electron-lattice equilibration ($\tau_{e-l}<\SI{1}{\pico\second}$~\cite{bonfiglio2021sub,banerjee2021ultrafast}). The system instead primarily relaxes under non-equilibrium via exchange-driven demagnetization, conserving the total angular momentum~\cite{Note2}. This directly leads to the implication that HI-AOS can only be achieved at starting temperatures below the compensation point, where the starting point ($S_\textrm{4a,0}$,$S_\textrm{4c,0}$) is above the dashed-dotted purple line (case [i] in Fig.~\ref{f:concept2}). As $T_{0}$ drops further below $T_\textrm{comp}$, $|S_{4c,0}|$ grows twice faster than $|S_{4a,0}|$, permitting increasingly longer pulses to achieve HI-AOS since the system can tolerate some loss of angular momentum to the lattice alongside the exchange relaxation. If the starting temperature instead shifts just below $T_{comp}$, only a femtosecond pulse can realise HI-AOS since the system must conserve all angular momentum in order to access the appropriate non-equilibrium state (blue manifold). At starting temperatures above $T_\textrm{comp}$ (case [ii] in Fig.~\ref{f:concept2}), however, pure exchange-driven dynamics becomes incapable of driving the system in to the prerequisite non-equilibrium state required for HI-AOS. This qualitatively explains why MRG displays HI-AOS only when $T_{0}<T_\textrm{comp}$ (Fig.~\ref{Davies2020_tau}).

\section{Outlook}\label{Outlook}

In this review, we have concentrated on qualitatively explaining the kinetics of HI-AOS found in ferrimagnetic GdFeCo and MRG alloys using a phenomenological framework. Since the first experimental identification of HI-AOS, more than ten distinct models have been constructed to account for the process~\cite{Ostler2012,Mentink2012,Wienholdt2013,Gridnev2018,Atxitia2013,Baral2015,Schellekens2013,Chimata2015,Barker2013,Iacocca2019,Mekonnen2013,Pelloux2020}. While these models all use different methods, they produce similar temporal dynamics of HI-AOS, and are qualitatively consistent with the phenomenological model presented here. Indeed, state-of-the-art atomistic simulations of HI-AOS also now embrace the same concepts of exchange and relativistic relaxations as used in the phenomenological approach~\cite{jakobs2022atomistic}. Quantitative differences between the approaches naturally emerge, but one must always be strongly guided by experimental measurements when studying the counter-intuitive non-equilibrium dynamics of ultrafast demagnetization that evidently underpin HI-AOS.

Despite the successes of the phenomenological model presented here, this model also exhibits limitations. For example, from the model it follows that the key ingredient to explain the counter-intuitive dynamics stems from exchange of angular momentum between different magnetic sublattices in the exchange-dominated regime. It is natural to expect that such a regime indeed exists for magnetic sublattices which themselves are weakly coupled to the environment. This situation may however not be realized in systems featuring magnetic ions with sizable orbital moments, such as for Tb ions. These are generally more strongly coupled to the lattice. It is possible to include single-ion anisotropy, even within the nonequilibrium free energy~\cite{mentink2012magnetism}, but this approach has so far not been investigated in the context of magnetization switching. Moreover, in systems with strong orbital moments, non-collinear effects may become important. The generalization of the current model to the non-collinear case has already been done for systems with weak coupling to the environment~\cite{bar2013exchange}, disclosing an additional exchange-driven pathway for precessional switching in ferrimagnets. In addition, the same equations were applied to describe the AFM-FM transition in FeRh systems~\cite{li2020ultrafast,dolgikh2022ultrafast}, which was found to yield profound differences between collinear and noncollinear phases. However, the combination of noncollinearity and sizeable orbital moments has not, however, been investigated so far. Moreover, being based on a macroscopic description of the sublattice magnetizations only, the phenomenological model fails in describing possible noncollinear magnetic states that can emerge after laser heating~\cite{berruto2018laser,je2018creation,buttner2021observation}. 

An open and outstanding question relates to whether the models referred to above, including the phenomenological model, can explain the HI-AOS experimentally identified in the synthetic ferrimagnets Gd/Co and Tb/Co multilayers, and TbCo alloys doped with minute amounts of Gd. Beens \textit{et al.} have successfully generalized the microscopic three-temperature model to account for the switching found in Gd/Co synthetic ferrimagnets~\cite{beens2019comparing}. There, the switching is primarily identified as an exchange-dominated effect with the reversal occurring close to the interface on the Co side, and subsequently propagating in to the bulk of the nanolayer. We anticipate that the phenomenological model presented here can reproduce this behavior, with appropriate division of the nanolayers in to even smaller thicknesses.

At the same time, it is not yet fully resolved why Gd facilitates HI-AOS whereas Tb does not. Early experiments studying fs-laser-induced ultrafast demagnetization in TbFeCo alloys identified the formation of a transient ferromagnetic-like state, with the magnetization of the Fe sublattice temporarily crossing zero~\cite{khorsand2013element}. Despite the ferromagnetic-like state persisting for more than \SI{25}{\pico\second}, the magnetization switched back. It was suggested that the strong spin-orbit coupling of the Tb sublattice provides an additional channel of angular momentum dissipation in competition with exchange-relaxation. Such a competing force is absent in Gd, which has an exactly half-filled 4$f$ shell producing zero net orbital moment. This interpretation is complicated however by recent experimental works showing that TbCo alloys doped with just 1.5\% of Gd display HI-AOS~\cite{zhang2022role}. Several research groups, using state-of-the-art atomistic models, currently argue that sublattice-specific damping plays a crucial role in the kinetics of HI-AOS~\cite{ceballos2021role,zhang2022role}.

The explanation of HI-AOS in synthetic ferrimagnets of Tb/Co nanostructures probably represents an even more difficult challenge. In the first experiments, Avil{\'e}s-F{\'e}lix \textit{et. al.} tested HI-AOS as a function of nanolayer thickness $t_\textrm{Co,Tb}$~\cite{aviles2019integration,aviles2020single,li2021thz} and found that the switching can only be achieved for thickness ratios $t_\textrm{Co}/t_\textrm{Tb}\leq1.2$ with $t_\textrm{Co}>\SI{1}{\nano\meter}$. As a point of comparison, the compensation point is at $\approx t_\textrm{Co}/t_\textrm{Tb}\leq1.1$. Switching could be achieved using either \SI{100}{\femto\second}- or \SI{5}{\pico\second}-long pulses but, for some specific combination of thicknesses, HI-AOS was only attainable using the \SI{5}{\pico\second}-long pulses, with the fs pulses only inducing demagnetization. This behavior has not been identified before in all prior demonstrations of HI-AOS, and indicates that the switching mechanism may be very different from the process found in GdFeCo and MRG. Furthermore, single-shot pump-probe microscopy measurements indicate that the dynamics of HI-AOS found in Tb/Co multilayers initially involves the emergence of a ring of switched magnetization which subsequently collapses to a single switched domain, stabilizing on a timescale of $\approx \SI{100}{\pico\second}$~\cite{mishra2022towards}. This behavior is not understood at the time of writing. 

A significant number of works have attempted already to formulate rules that give rise to HI-AOS in ferrimagnetic alloys~\cite{davies2020pathways,atxitia2015optimal,moreno2017conditions}, but this generalization was severely impeded in the past by the fact that, for almost a decade, only GdFeCo alloys were known to display HI-AOS. The recent discovery of HI-AOS in MRG has offered a much better grounding for such proposals, with an empirical examination of common features between GdFeCo and MRG indicating the important role of exchange couplings within the ferrimagnet. Specifically, the sublattices must be strongly coupled antiferromagnetically to allow for exchange-relaxation, which is the only mechanism by which the constituent spins can cross zero and change sign. At the same time, both GdFeCo and MRG have dissimilar intra-sublattice exchange couplings. This imbalance is vital for the realization of exchange-driven switching since the angular momentum of one sublattice must grow at the expense of the other~\cite{davies2020exchange,davies2020pathways}. Beyond this, however, any further generalization must still fully explain, for example, the role of spin-orbit coupling.

To address the aforementioned gaps in our understanding, further experimental campaigns studying the temporal dynamics of HI-AOS are essential. While technically very challenging, it might be expected that XMCD measurements of the HI-AOS displayed by Tb/Co multilayers, using x-rays tuned to probe the sublattice-specific response of Tb and Co, could provide the crucial insight required to understand why the switching occurs or not. This approach would further require spatial resolution, in order to overcome the apparent spatial non-uniformity of the magnetization dynamics~\cite{mishra2022towards}.

Despite the many outstanding questions relating to the general physics and kinetics of HI-AOS, tremendous progress has been made in rendering the process compatible with data storage and information processing technologies. HI-AOS can now be achieved, for example, using ultrafast electrical pulses that can arguably be delivered in contemporary integrated circuitry~\cite{polley2022progress}. Time-resolved measurements of HI-AOS in GdCo dots has revealed that reducing the dot's size towards the nanoscale actually enhances the speed of switching~\cite{el2019ultrafast}, with 75\% reversal being achieved within $\approx$\SI{2}{\pico\second}. Moreover, by growing GdCo on a silicon substrate, it is possible to switch magnetization repeatedly using two appropriately-tuned \SI{250}{\femto\second}-long pulses separated in time by just \SI{7}{\pico\second}, corresponding to a write/erase frequency of \SI{140}{\giga\hertz}~\cite{steinbach2022accelerating}. Finally, large strides are being made in developing all-optically-switchable perpendicular magnetic tunnel junctions~\cite{chen2017all,borisov2016tunnelling}, already yielding tunneling magnetoresistance signals in excess of 40\%~\cite{aviles2020single}. Data-recording technologies beyond the state-of-the-art will undoubtedly benefit not only from the rapid progress made in understanding HI-AOS but also from the emerging progress in finding alternative means to all-optically switch magnetization, particularly with reference to non-thermal ultrafast and minimally-dissipative methods based on the selective excitation of electronic or phononic resonances~\cite{stupakiewicz2017ultrafast,stupakiewicz2021ultrafast}.

\section*{Acknowledgements}

This review has only been made possible by the dedicated work of our co-workers D.~V.~Afanasiev, C.~Banerjee, J.~Besbas, G.~Bonfiglio, J.~M.~D.~Coey, O.~Eriksson, J.~Hellsvik, B.~A.~Ivanov, T.~Janssen, M.~I.~Katsnelson, A.~F.~G.~van~der~Meer, K.~Rode, P.~Stamenov, A.~Stupakiewicz and A.~Tsukamoto. We are also grateful to the skillful technical support provided by C. Berkhout, S. Semin, A. J. Toonen and the technical staff at FELIX. We are grateful to J.-Y.~Bigot, J.~Bokor, U.~Bovensiepen, R.~Chantrell, O.~Chubykalo-Fesenko, H.~D\"{u}rr, V.~N.~Gridnev, B.~Koopmans, S.~Mangin, U.~Nowak, T.~A.~Ostler, R.~V.~Pisarev, I.~Radu, as well as to all Ph.D. students and postdoctoral fellows of the Condensed Matter Physics, Ultrafast Spectroscopy of Correlated Materials and Spectroscopy of Solids and Interfaces groups for stimulating and fruitf discussions. This research has received funding from Nederlandse Organisatie voor Wetenschappelijk Onderzoek (NWO) and the European Research Council ERC grant agreement No. 856538 (3D-MAGiC), and is part of the Shell-NWO/FOM-initiative ``Computational sciences for energy research” of Shell and Chemical Sciences, Earth and Life Sciences, Physical Sciences, FOM and STW.

\bibliography{mybibfile}

\begin{thebibliography}{10}
\expandafter\ifx\csname url\endcsname\relax
  \def\url#1{\texttt{#1}}\fi
\expandafter\ifx\csname urlprefix\endcsname\relax\def\urlprefix{URL }\fi
\expandafter\ifx\csname href\endcsname\relax
  \def\href#1#2{#2} \def\path#1{#1}\fi

\bibitem{bean1959superparamagnetism}
C.~Bean, J.~D. Livingston, Superparamagnetism, Journal of Applied Physics
  30~(4) (1959) S120--S129.

\bibitem{richter2007transition}
H.~J. Richter, The transition from longitudinal to perpendicular recording,
  Journal of Physics D: Applied Physics 40~(9) (2007) R149.

\bibitem{kryder2008heat}
M.~H. Kryder, E.~C. Gage, T.~W. McDaniel, W.~A. Challener, R.~E. Rottmayer,
  G.~Ju, Y.-T. Hsia, M.~F. Erden, Heat assisted magnetic recording, Proceedings
  of the IEEE 96~(11) (2008) 1810--1835.

\bibitem{challener2009heat}
W.~A. Challener, C.~Peng, A.~V. Itagi, D.~Karns, W.~Peng, Y.~Peng, X.~Yang,
  X.~Zhu, N.~J. Gokemeijer, Y.-T. Hsia, et~al., Heat-assisted magnetic
  recording by a near-field transducer with efficient optical energy transfer,
  Nature Photonics 3~(4) (2009) 220--224.

\bibitem{zhu2007microwave}
J.-G. Zhu, X.~Zhu, Y.~Tang, Microwave assisted magnetic recording, IEEE
  Transactions on Magnetics 44~(1) (2007) 125--131.

\bibitem{okamoto2015microwave}
S.~Okamoto, N.~Kikuchi, M.~Furuta, O.~Kitakami, T.~Shimatsu, Microwave assisted
  magnetic recording technologies and related physics, Journal of Physics D:
  Applied Physics 48~(35) (2015) 353001.

\bibitem{Beaurepaire1996}
E.~Beaurepaire, J.-C. Merle, A.~Daunois, J.-Y. Bigot, Ultrafast spin dynamics
  in ferromagnetic nickel, Physical Review Letters 76~(22) (1996) 4250.

\bibitem{Stanciu2007}
C.~D. Stanciu, F.~Hansteen, A.~V. Kimel, A.~Kirilyuk, A.~Tsukamoto, A.~Itoh,
  T.~Rasing, {All-optical magnetic recording with circularly polarized light},
  {Physical Review Letters} {99}~({4}) ({2007}) {047601}.

\bibitem{Ostler2012}
T.~A. Ostler, J.~Barker, R.~F.~L. Evans, R.~W. Chantrell, U.~Atxitia,
  O.~Chubykalo-Fesenko, S.~El~Moussaoui, L.~B. P.~J. Le~Guyader, E.~Mengotti,
  L.~J. Heyderman, et~al., Ultrafast heating as a sufficient stimulus for
  magnetization reversal in a ferrimagnet, Nature Communications 3~(1) (2012)
  666.

\bibitem{Mentink2012}
J.~H. Mentink, J.~Hellsvik, D.~V. Afanasiev, B.~A. Ivanov, A.~Kirilyuk, A.~V.
  Kimel, O.~Eriksson, M.~I. Katsnelson, T.~Rasing, Ultrafast spin dynamics in
  multisublattice magnets, Physical Review Letters 108~(5) (2012) 057202.

\bibitem{kirilyuk2013laser}
A.~Kirilyuk, A.~V. Kimel, T.~Rasing, Laser-induced magnetization dynamics and
  reversal in ferrimagnetic alloys, Reports on Progress in Physics 76~(2)
  (2013) 026501.

\bibitem{el2020progress}
A.~El-Ghazaly, J.~Gorchon, R.~B. Wilson, A.~Pattabi, J.~Bokor, Progress towards
  ultrafast spintronics applications, Journal of Magnetism and Magnetic
  Materials 502 (2020) 166478.

\bibitem{polley2022progress}
D.~Polley, A.~Pattabi, J.~Chatterjee, S.~Mondal, K.~Jhuria, H.~Singh,
  J.~Gorchon, J.~Bokor, Progress toward picosecond on-chip magnetic memory,
  Applied Physics Letters 120~(14) (2022) 140501.

\bibitem{lambert2014all}
C.-H. Lambert, S.~Mangin, B.~C.~S. Varaprasad, Y.~Takahashi, M.~Hehn,
  M.~Cinchetti, G.~Malinowski, K.~Hono, Y.~Fainman, M.~Aeschlimann, et~al.,
  All-optical control of ferromagnetic thin films and nanostructures, Science
  345~(6202) (2014) 1337--1340.

\bibitem{mangin2014engineered}
S.~Mangin, M.~Gottwald, C.-H. Lambert, D.~Steil, V.~Uhl{\'\i}{\v{r}}, L.~Pang,
  M.~Hehn, S.~Alebrand, M.~Cinchetti, G.~Malinowski, et~al., Engineered
  materials for all-optical helicity-dependent magnetic switching, Nature
  Materials 13~(3) (2014) 286--292.

\bibitem{medapalli2017multiscale}
R.~Medapalli, D.~Afanasiev, D.~K. Kim, Y.~Quessab, S.~Manna, S.~A. Montoya,
  A.~Kirilyuk, T.~Rasing, A.~V. Kimel, E.~E. Fullerton, {Multiscale dynamics of
  helicity-dependent all-optical magnetization reversal in ferromagnetic Co/Pt
  multilayers}, Physical Review B 96~(22) (2017) 224421.

\bibitem{Yamada2022efficient}
K.~T. Yamada, A.~V. Kimel, K.~H. Prabhakara, S.~Ruta, T.~Li, F.~Ando, S.~Semin,
  T.~Ono, A.~Kirilyuk, T.~Rasing, {Efficient all-optical helicity dependent
  switching of spins in a Pt/Co/Pt film by a dual-pulse excitation}, Frontiers
  in Nanotechnology 4.

\bibitem{stupakiewicz2017ultrafast}
A.~Stupakiewicz, K.~Szerenos, D.~Afanasiev, A.~Kirilyuk, A.~V. Kimel, Ultrafast
  nonthermal photo-magnetic recording in a transparent medium, Nature
  542~(7639) (2017) 71--74.

\bibitem{stupakiewicz2019selection}
A.~Stupakiewicz, K.~Szerenos, M.~D. Davydova, K.~A. Zvezdin, A.~K. Zvezdin,
  A.~Kirilyuk, A.~V. Kimel, Selection rules for all-optical magnetic recording
  in iron garnet, Nature Communications 10~(1) (2019) 1--7.

\bibitem{frej2021all}
A.~Frej, A.~Maziewski, A.~Stupakiewicz, {All-optical magnetic recording in
  garnets using a single laser pulse at L-band telecom wavelengths}, Applied
  Physics Letters 118~(26) (2021) 262401.

\bibitem{stupakiewicz2021ultrafast}
A.~Stupakiewicz, C.~S. Davies, K.~Szerenos, D.~Afanasiev, K.~S. Rabinovich,
  A.~V. Boris, A.~Caviglia, A.~V. Kimel, A.~Kirilyuk, Ultrafast phononic
  switching of magnetization, Nature Physics 17~(4) (2021) 489--492.

\bibitem{Radu2011}
I.~Radu, K.~Vahaplar, C.~Stamm, T.~Kachel, N.~Pontius, H.~A. D{\"u}rr, T.~A.
  Ostler, J.~Barker, R.~F.~L. Evans, R.~W. Chantrell, A.~Tsukamoto, A.~Itoh,
  A.~Kirilyuk, T.~Rasing, A.~V. Kimel, Transient ferromagnetic-like state
  mediating ultrafast reversal of antiferromagnetically coupled spins, Nature
  472~(7342) (2011) 205.

\bibitem{el2019ultrafast}
A.~El-Ghazaly, B.~Tran, A.~Ceballos, C.-H. Lambert, A.~Pattabi, S.~Salahuddin,
  F.~Hellman, J.~Bokor, Ultrafast magnetization switching in nanoscale magnetic
  dots, Applied Physics Letters 114~(23) (2019) 232407.

\bibitem{ceballos2021role}
A.~Ceballos, A.~Pattabi, A.~El-Ghazaly, S.~Ruta, C.~P. Simon, R.~F.~L. Evans,
  T.~Ostler, R.~W. Chantrell, E.~Kennedy, M.~Scott, et~al., {Role of
  element-specific damping in ultrafast, helicity-independent, all-optical
  switching dynamics in amorphous (Gd,Tb)Co thin films}, Physical Review B
  103~(2) (2021) 024438.

\bibitem{zhang2022role}
W.~Zhang, J.~X. Lin, T.~X. Huang, G.~Malinowski, M.~Hehn, Y.~Xu, S.~Mangin,
  W.~Zhao, {Role of spin-lattice coupling in ultrafast demagnetization and all
  optical helicity-independent single-shot switching in
  Gd$_{1-x-y}$Tb$_y$Co$_x$ alloys}, Physical Review B 105~(5) (2022) 054410.

\bibitem{banerjee2020single}
C.~Banerjee, N.~Teichert, K.~E. Siewierska, Z.~Gercsi, G.~Y.~P. Atcheson,
  P.~Stamenov, K.~Rode, J.~M.~D. Coey, J.~Besbas, {Single pulse all-optical
  toggle switching of magnetization without gadolinium in the ferrimagnet
  Mn$_2$Ru$_x$Ga}, Nature Communications 11~(1) (2020) 1--6.

\bibitem{davies2020exchange}
C.~S. Davies, G.~Bonfiglio, K.~Rode, J.~Besbas, C.~Banerjee, P.~Stamenov,
  J.~M.~D. Coey, A.~V. Kimel, A.~Kirilyuk, Exchange-driven all-optical magnetic
  switching in compensated 3\textit{d} ferrimagnets, Physical Review Research
  2~(3) (2020) 032044.

\bibitem{evans2014ultrafast}
R.~F.~L. Evans, T.~A. Ostler, R.~W. Chantrell, I.~Radu, T.~Rasing, Ultrafast
  thermally induced magnetic switching in synthetic ferrimagnets, Applied
  Physics Letters 104~(8) (2014) 082410.

\bibitem{gerlach2017modeling}
S.~Gerlach, L.~Oroszlany, D.~Hinzke, S.~Sievering, S.~Wienholdt, L.~Szunyogh,
  U.~Nowak, Modeling ultrafast all-optical switching in synthetic ferrimagnets,
  Physical Review B 95~(22) (2017) 224435.

\bibitem{tsema2017laser}
Y.~Tsema, Laser induced magnetization dynamics and switching in multilayers,
  Ph.D. thesis, Radboud University Nijmegen (2017).

\bibitem{lalieu2017deterministic}
M.~L.~M. Lalieu, M.~J.~G. Peeters, S.~R.~R. Haenen, R.~Lavrijsen, B.~Koopmans,
  Deterministic all-optical switching of synthetic ferrimagnets using single
  femtosecond laser pulses, Physical Review B 96~(22) (2017) 220411.

\bibitem{beens2019comparing}
M.~Beens, M.~L.~M. Lalieu, A.~J.~M. Deenen, R.~A. Duine, B.~Koopmans, Comparing
  all-optical switching in synthetic-ferrimagnetic multilayers and alloys,
  Physical Review B 100~(22) (2019) 220409.

\bibitem{aviles2019integration}
L.~Avil{\'e}s-F{\'e}lix, L.~{\'A}lvaro-G{\'o}mez, G.~Li, C.~S. Davies,
  A.~Olivier, M.~Rubio-Roy, S.~Auffret, A.~Kirilyuk, A.~V. Kimel, T.~Rasing,
  et~al., {Integration of Tb/Co multilayers within optically switchable
  perpendicular magnetic tunnel junctions}, AIP Advances 9~(12) (2019) 125328.

\bibitem{aviles2020single}
L.~Avil{\'e}s-F{\'e}lix, A.~Olivier, G.~Li, C.~S. Davies,
  L.~{\'A}lvaro-G{\'o}mez, M.~Rubio-Roy, S.~Auffret, A.~Kirilyuk, A.~Kimel,
  T.~Rasing, et~al., {Single-shot all-optical switching of magnetization in
  Tb/Co multilayer-based electrodes}, Scientific Reports 10~(1) (2020) 1--8.

\bibitem{khorsand2012role}
A.~R. Khorsand, M.~Savoini, A.~Kirilyuk, A.~V. Kimel, A.~Tsukamoto, A.~Itoh,
  T.~Rasing, Role of magnetic circular dichroism in all-optical magnetic
  recording, Physical Review Letters 108~(12) (2012) 127205.

\bibitem{gorchon2016model}
J.~Gorchon, Y.~Yang, J.~Bokor, Model for multishot all-thermal all-optical
  switching in ferromagnets, Physical review B 94~(2) (2016) 020409.

\bibitem{betto2015site}
D.~Betto, N.~Thiyagarajah, Y.-C. Lau, C.~Piamonteze, M.-A. Arrio, P.~Stamenov,
  J.~M.~D. Coey, K.~Rode, {Site-specific magnetism of half-metallic
  Mn$_2$Ru$_x$Ga thin films determined by X-ray absorption spectroscopy},
  Physical Review B 91~(9) (2015) 094410.

\bibitem{vzic2016designing}
M.~{\v{Z}}ic, K.~Rode, N.~Thiyagarajah, Y.-C. Lau, D.~Betto, J.~M.~D. Coey,
  S.~Sanvito, K.~J. O'Shea, C.~A. Ferguson, D.~A. MacLaren, et~al., Designing a
  fully compensated half-metallic ferrimagnet, Physical Review B 93~(14) (2016)
  140202.

\bibitem{steil2011}
D.~Steil, S.~Alebrand, A.~Hassdenteufel, M.~Cinchetti, M.~Aeschlimann,
  All-optical magnetization recording by tailoring optical excitation
  parameters, Physical Review B 84~(22) (2011) 224408.

\bibitem{vahaplar2012all}
K.~Vahaplar, A.~M. Kalashnikova, A.~V. Kimel, S.~Gerlach, D.~Hinzke, U.~Nowak,
  R.~Chantrell, A.~Tsukamoto, A.~Itoh, A.~Kirilyuk, et~al., All-optical
  magnetization reversal by circularly polarized laser pulses: {E}xperiment and
  multiscale modeling, Physical Review B 85~(10) (2012) 104402.

\bibitem{gorchon2016role}
J.~Gorchon, R.~B. Wilson, Y.~Yang, A.~Pattabi, J.~Y. Chen, L.~He, J.~P. Wang,
  M.~Li, J.~Bokor, {Role of electron and phonon temperatures in the
  helicity-independent all-optical switching of GdFeCo}, Physical Review B
  94~(18) (2016) 184406.

\bibitem{davies2020pathways}
C.~S. Davies, T.~Janssen, J.~Mentink, A.~Tsukamoto, A.~V. Kimel, A.~F.~G.
  van~der Meer, A.~Stupakiewicz, A.~Kirilyuk, Pathways for single-shot
  all-optical switching of magnetization in ferrimagnets, Physical Review
  Applied 13~(2) (2020) 024064.

\bibitem{jakobs2021unifying}
F.~Jakobs, T.~A. Ostler, C.-H. Lambert, Y.~Yang, S.~Salahuddin, R.~B. Wilson,
  J.~Gorchon, J.~Bokor, U.~Atxitia, {Unifying femtosecond and picosecond
  single-pulse magnetic switching in Gd-Fe-Co}, Physical Review B 103~(10)
  (2021) 104422.

\bibitem{ostler2011crystallographically}
T.~A. Ostler, R.~F.~L. Evans, R.~W. Chantrell, U.~Atxitia,
  O.~Chubykalo-Fesenko, I.~Radu, R.~Abrudan, F.~Radu, A.~Tsukamoto, A.~Itoh,
  et~al., Crystallographically amorphous ferrimagnetic alloys: {C}omparing a
  localized atomistic spin model with experiments, Physical Review B 84~(2)
  (2011) 024407.

\bibitem{yang2017ultrafast}
Y.~Yang, R.~B. Wilson, J.~Gorchon, C.-H. Lambert, S.~Salahuddin, J.~Bokor,
  Ultrafast magnetization reversal by picosecond electrical pulses, Science
  Advances 3~(11) (2017) e1603117.

\bibitem{mentink2012magnetism}
J.~H. Mentink, Magnetism on the timescale of the exchange interaction:
  explanations and predictions, Ph.D. thesis, Radboud University Nijmegen
  (2012).

\bibitem{radu2015ultrafast}
I.~Radu, C.~Stamm, A.~Eschenlohr, F.~Radu, R.~Abrudan, K.~Vahaplar, T.~Kachel,
  N.~Pontius, R.~Mitzner, K.~Holldack, et~al., Ultrafast and distinct spin
  dynamics in magnetic alloys, Spin 5 (2015) 1550004.

\bibitem{onsager1931reciprocal}
L.~Onsager, {Reciprocal relations in irreversible processes. I.}, Physical
  Review 37~(4) (1931) 405.

\bibitem{iwata1983thermodynamical}
T.~Iwata, A thermodynamical approach to the irreversible magnetization in
  single-domain particles, Journal of Magnetism and Magnetic Materials 31
  (1983) 1013--1014.

\bibitem{iwata1986irreversible}
T.~Iwata, Irreversible magnetization in some ferromagnetic insulators, Journal
  of Magnetism and Magnetic Materials 59~(3-4) (1986) 215--220.

\bibitem{baryakhtar1984phenomenological}
V.~G. Baryakhtar, Phenomenological description of relaxation processes in
  magnets, Zhurnal \'Eksperimental'no\v l i Teoretichesko\v l Fiziki 87~(4)
  (1984) 1501--1508.

\bibitem{bar1985phenomenological}
V.~G. Bar'yakhtar, Phenomenological description of exchange relaxation
  processes in antiferromagnets, Low Temperature Physics 11~(11) (1985)
  1198--1205.

\bibitem{baryakhtar1998phenomenological}
V.~G. Baryakhtar, The phenomenological theory of relaxation processes in
  magnets, Frontiers in Magnetism of Reduced Dimension Systems (1998) 63--94.

\bibitem{landau1935theory}
L.~Landau, On the theory of the dispersion of magnetic permeability in
  ferromagnetic bodies, Physik. Z. Sowjetunion 8 (1935) 153--169.

\bibitem{abubrig2001}
O.~F. Abubrig, D.~Horvath, A.~Bobak, M.~Ja{\v{s}}{\v{c}}ur, {Mean-field
  solution of the mixed spin-1 and spin-32 Ising system with different
  single-ion anisotropies}, Physica A: Statistical Mechanics and Its
  Applications 296~(3-4) (2001) 437--450.

\bibitem{falk1970inequalities}
H.~Falk, {Inequalities of JW Gibbs}, American Journal of Physics 38~(7) (1970)
  858--869.

\bibitem{atxitia2014controlling}
U.~Atxitia, J.~Barker, R.~W. Chantrell, O.~Chubykalo-Fesenko, Controlling the
  polarity of the transient ferromagneticlike state in ferrimagnets, Physical
  Review B 89~(22) (2014) 224421.

\bibitem{brown1963thermal}
W.~F. Brown~Jr, Thermal fluctuations of a single-domain particle, Physical
  Review 130~(5) (1963) 1677.

\bibitem{kubo1970brownian}
R.~Kubo, N.~Hashitsume, Brownian motion of spins, Progress of Theoretical
  Physics Supplement 46 (1970) 210--220.

\bibitem{garanin1997fokker}
D.~A. Garanin, {Fokker-Planck and Landau-Lifshitz-Bloch equations for classical
  ferromagnets}, Physical Review B 55~(5) (1997) 3050.

\bibitem{Note1}
For GdFeCo, we adopt the same magnetic parameters as used in
  Ref.~\cite{davies2020pathways} to model Gd$_{24}$Fe$_{76}$. For MRG, we fix
  the magnetic moments $\sigma_{4a}=$\SI{2}{\micro_B} and
  $\sigma_{4c}=$\SI{2.4}{\micro_B} and g-factors $g_{4a}\approx g_{4c}=2$,
  drawing from the results of x-ray MCD measurements~\cite{betto2015site} and
  density-functional calculations~\cite{vzic2016designing}. We also assume
  $J_{4a-4a}=$\SI{150}{\milli\electronvolt},
  $J_{4a-4c}=$\SI{-50}{\milli\electronvolt} and
  $J_{4c-4c}=$\SI{35}{\milli\electronvolt}, using the same typical ratio as
  found for Mn$_{2}$Ru$_{0.61}$Ga in Ref.~\cite{borisov2016tunnelling}.

\bibitem{koopmans2010explaining}
B.~Koopmans, G.~Malinowski, F.~Dalla~Longa, D.~Steiauf, M.~F{\"a}hnle, T.~Roth,
  M.~Cinchetti, M.~Aeschlimann, Explaining the paradoxical diversity of
  ultrafast laser-induced demagnetization, Nature Materials 9~(3) (2010)
  259--265.

\bibitem{bergeard2014ultrafast}
N.~Bergeard, V.~L{\'o}pez-Flores, V.~Halte, M.~Hehn, C.~Stamm, N.~Pontius,
  E.~Beaurepaire, C.~Boeglin, Ultrafast angular momentum transfer in
  multisublattice ferrimagnets, Nature Communications 5~(1) (2014) 1--7.

\bibitem{graves2013nanoscale}
C.~E. Graves, A.~H. Reid, T.~Wang, B.~Wu, S.~De~Jong, K.~Vahaplar, I.~Radu,
  D.~P. Bernstein, M.~Messerschmidt, L.~M{\"u}ller, et~al., {Nanoscale spin
  reversal by non-local angular momentum transfer following ultrafast laser
  excitation in ferrimagnetic GdFeCo}, Nature Materials 12~(4) (2013) 293--298.

\bibitem{bonfiglio2021sub}
G.~Bonfiglio, K.~Rode, G.~Y.~P. Atcheson, P.~Stamenov, J.~M.~D. Coey, A.~V.
  Kimel, T.~Rasing, A.~Kirilyuk, {Sub-picosecond exchange--relaxation in the
  compensated ferrimagnet Mn$_2$Ru$_x$Ga}, Journal of Physics: Condensed Matter
  33~(13) (2021) 135804.

\bibitem{banerjee2021ultrafast}
C.~Banerjee, K.~Rode, G.~Atcheson, S.~Lenne, P.~Stamenov, J.~M.~D. Coey,
  J.~Besbas, Ultrafast double pulse all-optical reswitching of a ferrimagnet,
  Physical Review Letters 126~(17) (2021) 177202.

\bibitem{Note2}
We note that the strength of the inter-sublattice exchange coupling in MRG,
  combined with the similarity of the sublattice-specific magnetic moments,
  would result in decoupled Bloch relaxations [following Eq.~(\ref{e:taui})]
  that is difficult to distinguish from exchange-driven
  relaxation~\cite{bonfiglio2021sub,banerjee2021ultrafast,jakobs2022atomistic},
  assuming similar relativistic relaxation constants. The fact that HI-AOS has
  not been experimentally found thus far in MRG samples at
  $T_{0}>T_\textrm{comp}$ suggests that decoupled Bloch relaxation does not
  dominate~\cite{banerjee2020single,davies2020exchange}. The exchange-driven
  relaxation is compulsory, of course, for entering the transient
  ferromagnetic-like state and thus achieving HI-AOS.

\bibitem{Wienholdt2013}
S.~Wienholdt, D.~Hinzke, K.~Carva, P.~M. Oppeneer, U.~Nowak, Orbital-resolved
  spin model for thermal magnetization switching in rare-earth-based
  ferrimagnets, Physical Review B 88~(2) (2013) 020406.

\bibitem{Gridnev2018}
V.~N. Gridnev, Ferromagneticlike states and all-optical magnetization switching
  in ferrimagnets, Physical Review B 98~(1) (2018) 014427.

\bibitem{Atxitia2013}
U.~Atxitia, T.~Ostler, J.~Barker, R.~F.~L. Evans, R.~W. Chantrell,
  O.~Chubykalo-Fesenko, Ultrafast dynamical path for the switching of a
  ferrimagnet after femtosecond heating, Physical Review B 87~(22) (2013)
  224417.

\bibitem{Baral2015}
A.~Baral, H.~C. Schneider, {Magnetic switching dynamics due to ultrafast
  exchange scattering: A model study}, Physical Review B 91~(10) (2015) 100402.

\bibitem{Schellekens2013}
A.~J. Schellekens, B.~Koopmans, Comparing ultrafast demagnetization rates
  between competing models for finite temperature magnetism, Physical Review
  Letters 110~(21) (2013) 217204.

\bibitem{Chimata2015}
R.~Chimata, L.~Isaeva, K.~K{\'a}das, A.~Bergman, B.~Sanyal, J.~H. Mentink,
  M.~I. Katsnelson, T.~Rasing, A.~Kirilyuk, A.~V. Kimel, et~al., All-thermal
  switching of amorphous {Gd-Fe} alloys: {A}nalysis of structural properties
  and magnetization dynamics, Physical Review B 92~(9) (2015) 094411.

\bibitem{Barker2013}
J.~F.~L. Barker, U.~Atxitia, T.~A. Ostler, O.~Hovorka, O.~Chubykalo-Fesenko,
  R.~W. Chantrell, Two-magnon bound state causes ultrafast thermally induced
  magnetisation switching, Scientific Reports 3 (2013) 3262.

\bibitem{Iacocca2019}
E.~Iacocca, T.-M. Liu, A.~H. Reid, Z.~Fu, S.~Ruta, P.~W. Granitzka, E.~Jal,
  S.~Bonetti, A.~X. Gray, C.~E. Graves, et~al., Spin-current-mediated rapid
  magnon localisation and coalescence after ultrafast optical pumping of
  ferrimagnetic alloys, Nature Communications 10~(1) (2019) 1756.

\bibitem{Mekonnen2013}
A.~Mekonnen, A.~Khorsand, M.~Cormier, A.~V. Kimel, A.~Kirilyuk, A.~Hrabec,
  L.~Ranno, A.~Tsukamoto, A.~Itoh, T.~Rasing, Role of the inter-sublattice
  exchange coupling in short-laser-pulse-induced demagnetization dynamics of
  {GdCo} and {GdCoFe} alloys, Physical Review B 87~(18) (2013) 180406.

\bibitem{Pelloux2020}
J.~Pelloux-Prayer, F.~Moradi, Compact model of all-optical-switching magnetic
  elements, IEEE Transactions on Electron Devices 67~(7) (2020) 2960--2965.

\bibitem{jakobs2022atomistic}
F.~Jakobs, U.~Atxitia, {Atomistic spin model of single pulse toggle switching
  in Mn$_2$Ru$_x$Ga Heusler alloys}, Applied Physics Letters 120~(17) (2022)
  172401.

\bibitem{bar2013exchange}
V.~G. Bar’yakhtar, V.~I. Butrim, B.~A. Ivanov, Exchange relaxation as a
  mechanism of the ultrafast reorientation of spins in a two-sublattice
  ferrimagnet, JETP letters 98~(5) (2013) 289--293.

\bibitem{li2020ultrafast}
G.~Li, R.~Medapalli, J.~H. Mentink, R.~V. Mikhaylovskiy, T.~G.~H. Blank,
  S.~K.~K. Patel, A.~K. Zvezdin, T.~Rasing, E.~E. Fullerton, A.~V. Kimel,
  {Ultrafast kinetics of the antiferromagnetic-ferromagnetic phase transition
  in FeRh}, arXiv preprint arXiv:2001.06799.

\bibitem{dolgikh2022ultrafast}
I.~A. Dolgikh, T.~G.~H. Blank, G.~Li, K.~H. Prabhakara, S.~K.~K. Patel, A.~G.
  Buzdakov, R.~Medapalli, E.~E. Fullerton, O.~V. Koplak, J.~H. Mentink, et~al.,
  {Ultrafast emergence of ferromagnetism in antiferromagnetic FeRh in high
  magnetic fields}, arXiv preprint arXiv:2202.03931.

\bibitem{berruto2018laser}
G.~Berruto, I.~Madan, Y.~Murooka, G.~M. Vanacore, E.~Pomarico, J.~Rajeswari,
  R.~Lamb, P.~Huang, A.~J. Kruchkov, Y.~Togawa, et~al., {Laser-induced skyrmion
  writing and erasing in an ultrafast cryo-Lorentz transmission electron
  microscope}, Physical Review Letters 120~(11) (2018) 117201.

\bibitem{je2018creation}
S.-G. Je, P.~Vallobra, T.~Srivastava, J.-C. Rojas-S{\'a}nchez, T.~H. Pham,
  M.~Hehn, G.~Malinowski, C.~Baraduc, S.~Auffret, G.~Gaudin, et~al., Creation
  of magnetic skyrmion bubble lattices by ultrafast laser in ultrathin films,
  Nano Letters 18~(11) (2018) 7362--7371.

\bibitem{buttner2021observation}
F.~B{\"u}ttner, B.~Pfau, M.~B{\"o}ttcher, M.~Schneider, G.~Mercurio, C.~M.
  G{\"u}nther, P.~Hessing, C.~Klose, A.~Wittmann, K.~Gerlinger, et~al.,
  Observation of fluctuation-mediated picosecond nucleation of a topological
  phase, Nature Materials 20~(1) (2021) 30--37.

\bibitem{khorsand2013element}
A.~R. Khorsand, M.~Savoini, A.~Kirilyuk, A.~V. Kimel, A.~Tsukamoto, A.~Itoh,
  T.~Rasing, Element-specific probing of ultrafast spin dynamics in
  multisublattice magnets with visible light, Physical Review Letters 110~(10)
  (2013) 107205.

\bibitem{li2021thz}
G.~Li, Thz spintronics at interfaces of metals, Ph.D. thesis, Radboud
  University Nijmegen (2021).

\bibitem{mishra2022towards}
K.~G. Mishra, Towards nanoscale confinement of all optical magnetization
  switching, Ph.D. thesis, Radboud University Nijmegen (2022).

\bibitem{atxitia2015optimal}
U.~Atxitia, T.~A. Ostler, R.~W. Chantrell, O.~Chubykalo-Fesenko, Optimal
  electron, phonon, and magnetic characteristics for low energy thermally
  induced magnetization switching, Applied Physics Letters 107~(19) (2015)
  192402.

\bibitem{moreno2017conditions}
R.~Moreno, T.~Ostler, R.~Chantrell, O.~Chubykalo-Fesenko, {Conditions for
  thermally induced all-optical switching in ferrimagnetic alloys: Modeling of
  TbCo}, Physical Review B 96~(1) (2017) 014409.

\bibitem{steinbach2022accelerating}
F.~Steinbach, N.~Stetzuhn, D.~Engel, U.~Atxitia, C.~von Korff~Schmising,
  S.~Eisebitt, Accelerating double pulse all-optical write/erase cycles in
  metallic ferrimagnets, Applied Physics Letters 120~(11) (2022) 112406.

\bibitem{chen2017all}
J.-Y. Chen, L.~He, J.-P. Wang, M.~Li, All-optical switching of magnetic tunnel
  junctions with single subpicosecond laser pulses, Physical Review Applied
  7~(2) (2017) 021001.

\bibitem{borisov2016tunnelling}
K.~Borisov, D.~Betto, Y.-C. Lau, C.~Fowley, A.~Titova, N.~Thiyagarajah,
  G.~Atcheson, J.~Lindner, A.~M. Deac, J.~M.~D. Coey, et~al., {Tunnelling
  magnetoresistance of the half-metallic compensated ferrimagnet
  Mn$_2$Ru$_x$Ga}, Applied Physics Letters 108~(19) (2016) 192407.

\end{thebibliography}

\end{document}